\documentclass[11pt]{article}

\newcommand{\hiddenforspeed}[1]{#1}
    
\usepackage{mathtools} 
\usepackage{booktabs} 
\usepackage{tikz} 

\usepackage{amsthm}
\usepackage{pgfplots}
\pgfplotsset{compat=1.15}
\usepackage{todonotes} 
\usepackage{subcaption}
\usepackage{amsfonts}
\usepackage{nicefrac}
\usepackage{mathtools}
\usepackage{wrapfig}
\usepackage[inline]{enumitem}

\usepackage[square,numbers,sort&compress]{natbib}
\usepackage[margin=1in]{geometry}

\newcommand{\removememaybe}[1]{}
\usepackage{authblk}

\usepackage[inline]{enumitem}

\usepackage[T1]{fontenc}

\newcommand{\Winner}{\textsc{Winner}}
\newcommand{\bordaOWAwinner}[1][]{%
	$\beta$%
	\ifx&#1&%
	\else
	(#1)%
	\fi%
	-\Winner%
}%

\allowdisplaybreaks

\clearpage{}%

\newcommand{\normphi}{{{\mathrm{norm}\hbox{-}\phi}}}

\newcommand{\first}{}    
\def\first/{red}
\newcommand{\second}{}    
\def\second/{blue}
\newcommand{\third}{}    
\def\third/{black}
\newcommand{\fourth}{}    
\def\fourth/{green}
\clearpage{}%
\usepackage{hyperref}
\usepackage{cleveref}

\usepackage{siunitx}
 
 \title{A Quantitative and Qualitative Analysis of the Robustness of (Real-World) Election Winners}
 \author[1]{Niclas Boehmer}
 \author[2]{Robert Bredereck}
 \author[3]{Piotr Faliszewski}
 \author[1]{Rolf Niedermeier}
 \affil[1]{\small
 	Algorithmics and Computational 
 	Complexity, Technische Universit\"at Berlin, Berlin,\protect\\
 	niclas.boehmer@tu-berlin.de}
 \affil[2]{\small
 	TU Clausthal, Clausthal-Zellerfeld,
 	robert.bredereck@tu-clausthal.de}
  \affil[3]{\small
 	AGH University, Krak\'{o}w,
 	faliszew@agh.edu.pl}
 \date{\today}
 
\begin{document}

\maketitle
 
\begin{abstract}
    Contributing to the toolbox for interpreting election results, we evaluate the robustness of election winners to random noise. 
    We compare the robustness of different voting rules and
    evaluate the robustness of real-world election winners from the Formula 1 World Championship and some variant of political elections.
    We find many instances of elections  that have very non-robust winners and numerous delicate robustness patterns that cannot be identified using classical and simpler approaches.
\end{abstract}

\maketitle

\section{Introduction}\label{sec:intro}
Voting is a convenient and powerful framework to aggregate preferences. 
It has many real-world applications ranging from political elections, through evaluation panels deciding on which research projects
 to fund, and televoting in TV shows, to aggregating the results of sport competitions. 
Interestingly, independent of the application, one regularly, and emotionally, debated matter is by ``how much'' the winning candidate had won the election. 
Remarkably, studies have found that there are more extremely close elections than one might intuitively expect. 
For instance, there is a list of $313$ political elections on Wikipedia where the election was decided by less than $0.1\%$ of all voters \cite{wiki:close}.
Moreover, \citet{mulligan2003empirical} reported that in state elections in the United States one in every $\num{15,000}$ voters casts a decisive vote. 
Motivated by this, there is a rich body of theoretical literature on the likelihood that elections are decided by a single vote \cite{DBLP:conf/sigecom/Xia21,banzhaf1968one,beck1975note,margolis1977probability,chamberlain1981note,gillett1977collective,gillett1980comparative,marchant2001probability}.

But what is the relevance of close elections beyond being a topic people like to argue about? 
The main underlying assumption here is that the recorded votes in the election capture reality only approximately. 
For instance, it might be the case that some voters cast their votes in a rush or without having enough information available to them, voters were unable to participate in the election, or votes were incorrectly recorded due to technical errors (which indeed happen sometimes \cite{norden2007post,wolchok2010security}).
If an election is detected to be close, various counter-measures can be taken: 
For instance, it is possible to do a recount or to audit the election results \cite{stark2008conservative,stark2008sharper,stark2009efficient,sarwate2011risk}, to continue discussions about the election issue, or to collect further votes.
Even if one assumes that the election result is ``correct'', by how much a candidate won also influences its legitimacy and credibility, in particular considering that voters might change their mind over time. 
To sum up, a reliable estimate for the lead of an election winner has the potential to increase the fairness and transparency of elections, as it allows for a better interpretation of election results and the initiation of possible countermeasures. 
 
But what does it mean for an election to be close? 
In political elections, Plurality voting is often used. Here, each voter awards one point to its most preferred candidate and the candidate with the most points wins.
For Plurality (and also for arbitrary scoring-based rules) a natural and common measure to assess the closeness of an election is the difference between the score of the election winner and the candidate finishing in the second place. 
A more fine grained version of this notion is the \emph{margin of victory}, which is defined as the minimum number of voters that need to change their votes to change the election outcome \cite{DBLP:conf/ijcai/DeyN15,DBLP:conf/uss/MagrinoRS11,DBLP:conf/uss/Cary11,DBLP:conf/sigecom/Xia12}.
However, both of these concepts are too ``coarse'' in certain situations.
To illustrate this, consider an election $E$ containing $50$ times the vote  $a\succ b\succ \dots$ and $49$ times the vote $b\succ \dots \succ a$, and an election $E'$ containing $50$ times the vote $a\succ \dots \succ b$ and $49$ times the vote $b\succ a\succ \dots$ (we write $a\succ b$ to indicate that $a$ is preferred to $b$ and ``$\dots$'' to indicate that we rank all remaining candidates in some arbitrary ordering).
While in both elections the score difference and margin of victory under Plurality is one, examining the votes more closely, the situation in these two elections is quite different: 
In $E$, in order for $b$ to win the election, only one of $50$ voters needs to slightly change its mind (by swapping its two most preferred candidates).
In contrast, in $E'$, in order for $b$ to win the election, at least one of $50$ voters needs to drastically change its mind by ranking its previously last-ranked candidate in the first position (plus there are $49$ voters where $a$ can easily gain a point).
To sum up, both the score difference and the margin of victory do not take into account that small changes are more likely than large ones. 
Further, both measures suffer from the drawback that they focus on the worst case (e.g., for the margin of victory it does not make a difference whether there is only one specific voter that can change the election outcome by modifying its vote or whether multiple voters have this power) and not on the average case, which is probably the practically more relevant one. 

In this paper, following the works of \citet{DBLP:conf/ijcai/BoehmerBFN21} and \citet{DBLP:conf/eumas/BaumeisterH21s},  we study a more fine-grained robustness measure:
We analyze how the winning probabilities of candidates behave if we start to perturb the election by performing some swaps of adjacent candidates in some votes.
How quickly the winning probability of the original election winner decreases as we move further and further away from the original election sheds some light on this winner's robustness.
Herein, we assume that changes are equiprobable and, thus, affect each voter and each part of the vote with the same probability. 
Moreover, we assume that the probability of replacing a vote by a new one is anti-proportional to the swap distance between the two.
For this, we make use of the famous Mallows noise model \cite{mallows1957non} (see \Cref{sec:assessing} for a detailed description of our approach). 
One can thus interpret our approach as a tool to measure the robustness of election winners against random equiprobable noise.

Recently, \citet{DBLP:conf/ijcai/BoehmerBFN21} conducted some experiments on the winning probabilities of candidates under Plurality and Borda on synthetic elections if elections are perturbed. 
Their results indicate that some elections have extremely non-robust winners and that a winning-probability based approach offers a different and more nuanced view on the robustness of election winners than established, simpler measures. 
In this work, we aim for  comparing the robustness of different voting rules and conducting an in-depth analysis of the robustness of winners in real-world elections.

\subsection{Related Work}
Closest works related to ours are the papers of \citet{DBLP:conf/eumas/BaumeisterH21s} and \citet{DBLP:conf/ijcai/BoehmerBFN21}.
Both study the computational complexity of computing the winning probabilities of candidates if we replace each vote with one sampled from some distribution. 
Among other models, \citet{DBLP:conf/eumas/BaumeisterH21s} considered a Mallows-based approach as used in this paper from a theoretical perspective: 
For some given $\phi\in [0,1]$, each vote $v$ is replaced by a vote sampled from a Mallows distribution with center vote $v$ and dispersion parameter $\phi$, i.e., a vote $v'$ is sampled with probability proportional to $\phi^{\kappa(v,v')}$ (the swap distance $\kappa(v,v')$ between~$v$ and~$v'$ is the number of swaps of adjacent candidates that are needed to transform~$v$ into~$v'$). 
\citet{DBLP:conf/ijcai/BoehmerBFN21} studied the related computational problem of counting elections at some given swap distance from a given initial election where some given candidate wins (see \Cref{sec:assessing} for details). 
In fact, \citet{DBLP:conf/eumas/BaumeisterH21s} showed these two problems to be equivalent from the computational perspective.
Together, \citet{DBLP:conf/eumas/BaumeisterH21s} and \citet{DBLP:conf/ijcai/BoehmerBFN21} proved strong (parameterized) intractability results for these problems for Plurality and Borda. 

The problem of \citet{DBLP:conf/ijcai/BoehmerBFN21} can be  phrased as the counting variant of the \textsc{Swap Bribery} problem. 
In \textsc{Swap Bribery}, we are given an election, a designated candidate $p$, and a budget $k$, and the question is whether we can perform $k$ swaps of adjacent candidates in some votes to make $p$ an election winner. 
Bribery problems in elections have been introduced by \citet{DBLP:journals/jair/FaliszewskiHH09} and have been extensively studied since then (see the overview of \citet{DBLP:reference/choice/FaliszewskiR16}). 
The idea to use swap bribery for evaluating the robustness of election winners is due to \citet{DBLP:conf/atal/ShiryaevYE13} (and has also been used in other contexts \cite{DBLP:journals/ai/BrillSS22,DBLP:journals/jair/BoehmerBHN21,DBLP:conf/ijcai/BoehmerBKL20}): In the \textsc{Destructive Swap Bribery} problem we want to prevent a given candidate from winning the election by performing as few swaps as possible. 
The minimum cost of a successful destructive swap bribery can then act as a robustness measure (however, like the margin of victory and score difference, this measure is focused on the worst case). 
\citet{DBLP:conf/ijcai/BoehmerBFN21} observed that for Borda and Plurality the minimum cost of a destructive swap bribery might be disconnected from their winning-probability based approach (see \Cref{sec:assessing}). 

\subsection{Our Contributions}

The main goal of this paper is to better understand the robustness of election winners against random equiprobable noise.
In particular, we analyze what makes an election winner robust and how this is influenced by the voting rule used. 
We address this goal in multiple steps, thereby significantly extending the experimental work of \citet{DBLP:conf/ijcai/BoehmerBFN21}, who only considered the robustness of winners under the Plurality and Borda voting rules in synthetic elections:
In \Cref{sec:assessing}, we present our approach for measuring the robustness of election winners and compare it to the approach used by \citet{DBLP:conf/ijcai/BoehmerBFN21}.  In essence, our measure
    is very similar but easier to handle and compute.
In \Cref{sub:comp-mal}, we compare the robustness of different voting rules on synthetic data. Generally speaking, out of the considered rules, Copeland tends to produce the most robust winners, then comes Borda, then Bucklin, then STV, and Plurality produces the least robust winners. 
In \Cref{se:real-world-exp}, we analyze the robustness of real-world elections from two different sources, i.e., the Formula 1 World Championship and some form of political elections. 
    We identify many elections with winners that are remarkably sensitive to random equiprobable noise. For example, in some editions of the Formula 1 World Championship the original winner loses with $22\%$ probability if we make only an expected number of $5$ \emph{random} swaps of adjacent candidates in the whole election. 
    
Furthermore, throughout the whole paper, we observe in different places that our approach allows one to identify patterns that are invisible when considering simpler robustness measures such as the score difference, and that the non-robustness of winners can be of different types.
Moreover, we describe how our approach can be used to distinguish between tied election winners, thereby serving as a potential tie-breaking mechanism. 

\section{Preliminaries}

\paragraph{Elections}
An election is a pair $(C,V)$ where $C=\{c_1,\dots, c_m\}$ is a set of candidates and $V=(v_1,\dots, v_n)$ is a collection of votes. 
Each vote is a strict total order over all candidates.
We write $v:c_1\succ c_2$ to denote that $v$ prefers $c_1$ to $c_2$, and for a candidate $c\in C$ we say that $v$ ranks $c$ in the $i$th position if $v$ prefers exactly $i-1$  candidates to $c$. 
In \Cref{se:real-world-exp}, we allow for top-truncated votes, i.e., strict total orders over subsets of candidates.
The implicit meaning of a top-truncated vote is that the voter prefers all the ranked candidates to all the unranked ones.
For a top-truncated vote, we refer to the number of candidates the voter ranks as the vote length. 

\paragraph{Voting Rules}
A voting rule is a function that maps an election to a subset of candidates that tie as winners of this election.
A \emph{scoring vector} is a vector $\mathbf{s}=(s_1,\dots, s_m)$ with $s_i\in \mathbb{R}$ for all $i\in [m]$ and $s_1\geq s_2\geq \dots \geq s_m$. A \emph{positional scoring rule} is defined by a scoring vector $\mathbf{s}$: Each voter awards $s_i$ points to the candidate it ranks in the $i$th position for each $i\in [m]$.\footnote{For top-truncated votes in an election with $m$ candidates, we still use the original scoring vector containing $m$ entries. A voter which ranks $j\in [m]$ candidates then awards $s_i$ points to the candidate it ranks in the $i$th position for each $i\in [j]$.}
All candidates with the maximum summed score win. 
Two rules that are of particular importance in our analysis are Plurality, which corresponds to the  scoring vector $(1,0,\dots,0)$, and Borda, which corresponds to the scoring vector $(m-1,m-2,\dots,1,0)$.  

Under the \emph{Copeland} voting rule, we compute a score for each candidate and all candidates with the highest score win. A candidate $c$ gets a point for each candidate $d\in C\setminus \{c\}$ for which more than half of the voters prefers $c$ to $d$ and loses a point for each candidate $d\in C\setminus \{c\}$ where more than half of the voters prefers $d$ to~$c$.

Under the \emph{Bucklin} voting rule, for each candidate let $i_c$ be the minimum $i\in [m]$ such that more than half of the voters rank $c$ in one of the first $i$ positions.
The candidate for which $i_c$ is smallest wins.  If multiple candidates have the minimum $i_c$, say~$i^*$, then the candidate(s) that appear in the most votes in one of the first $i^*$ positions win. 

Under the \emph{single transferable vote} (STV) voting rule, we are given a strict total order $\succ_t$ of the candidates as the tie-breaking order. 
In each round, we delete the candidate with the minimum Plurality score. 
If multiple candidates have the minimum Plurality score, then the candidate that is ranked last in $\succ_t$ is deleted. 
The last remaining candidate is the winner of the election.
Because we actively apply a tie-breaking rule for STV, there are no tied winners under STV.

\paragraph{Swap Distance} Given two votes $v$ and $v'$ over the same candidate set, their swap distance $\kappa(v,v')$ is the number of candidate pairs on whose ordering $v$ and $v'$ disagree (equivalently, this is the minimum number of swaps of adjacent candidates that are needed to transform $v$ into $v'$). 
Note that the maximum swap distance between two votes over $m$ candidates is $\frac{m(m-1)}{2}$.
The swap distance between two elections $E=(C,V)$ and $E'=(C,V')$ where $V=(v_1,\dots, v_n)$ and $V'=(v'_1,\dots, v'_n)$ is $\sum_{i=1}^{n} \kappa(v_i,v'_i)$. 

\paragraph{(Normalized) Mallows Distribution} 
For a set $C$ of $m$ candidates, the Mallows distribution \cite{mallows1957non} is parameterized by a central strict total order $v^*$ over $C$ and a dispersion parameter $\phi\in [0,1]$. 
It assigns to each strict total order $v$ over $C$ a probability $\mathcal{D}_{\text{Mallows}}^{v^*,\phi}(v)$ that depends on the swap distance between $v$ and $v^*$. 
Specifically, we have: 
$\mathcal{D}_{\text{Mallows}}^{v^{*},\phi}(v)=\frac{1}{Z}\phi^{\kappa(v,v^*)}$
with normalizing constant $Z=1\cdot (1+\phi)\cdot (1+\phi+\phi^2)\cdot \dots
\cdot(1+\dots+\phi^{m-1})$.
For $\phi=0$, vote $v^*$ has probability one and all other voters have probability zero. 
For $\phi=1$, all votes are drawn with the same probability. 
Note that we use the Mallows distribution in two different ways. 
On the one hand,  as part of our robustness measure, we use it to perturb a given vote $v'$, which typically means that we replace $v'$ by a vote sampled from $\mathcal{D}_{\text{Mallows}}^{v',\phi}$. 
On the other hand, we use it as a model to generate elections in which case we create an election by drawing multiple votes from $\mathcal{D}_{\text{Mallows}}^{v^{*},\phi}$ where $v^*$ is the lexicographic ordering of candidates.

Unfortunately, as argued by \citet{DBLP:conf/ijcai/BoehmerBFNS21}, the dispersion parameter $\phi$ is not easy to interpret. 
Moreover, elections with different numbers of candidates sampled from Mallows distributions with the same fixed value of $\phi$  are in some sense of a fundamentally different nature. 
That is why we use the normalization of Mallows model proposed by \citet{DBLP:conf/ijcai/BoehmerBFNS21}: 
Here, the Mallows distribution is parameterized by a normalized dispersion parameter $\normphi\in [0,1]$, which is internally converted to a value of~$\phi$, such that the expected swap distance between a sampled vote and the central vote is $\normphi\cdot \frac{m(m-1)}{4}$. 
Again, $\normphi=0$ leads to $v^*$ being sampled all the time
(the expected swap distance is zero)
and for $\normphi=1$ all votes have the same probability
(the expected swap distance is $\frac{m(m-1)}{4}$).
However, here, $\mathrm{norm}\hbox{-}\phi=0.5$ leads to a distribution that is in some sense in the middle between these two extremes, as the expected swap distance between the sampled and central vote is $\frac{m(m-1)}{8}$. 
Moreover, one value of $\normphi$ leads to the same expected relative number of  swaps for different numbers of candidates, which will be vital for our purposes.

Thus, using $\normphi$ instead of $\phi$ basically leads to a rescaling of the considered range of perturbation. 
That is, there is a one-to-one mapping of values of $\normphi$ and values of $\phi$.
However, as argued above, $\normphi$ is much easier to use and allows for a more natural interpretation and comparison of results.

\paragraph{Pearson Correlation Coefficient} The Pearson correlation coefficient (PCC) is a measure of a linear correlation between two quantities, where $1$ means perfect linear proportional correlation, $0$ means no linear correlation and $-1$ means  perfect linear anti-proportional correlation.

\section{Assessing Winner Robustness} \label{sec:assessing}

In this section, we describe how we asses the robustness of election winners by computing the candidate's probabilities to win the election if voters partly and randomly change their preferences.
In this section, to validate our approach, we briefly mention the results of some experiments which we conducted on a diverse collection of $800$ synthetic elections with $100$ voters and $10$ candidates collected by \citet{DBLP:conf/atal/SzufaFSST20} (this dataset has also been also used by \citet{DBLP:conf/ijcai/BoehmerBFN21}).

\begin{figure}
	\centering  
	\begin{minipage}[b]{0.58\textwidth}
		\begin{subfigure}{0.45\textwidth}
			\centering
			\resizebox{0.9\textwidth}{!}{\begin{tikzpicture}
\begin{axis}[anchor=north, no markers, enlargelimits=false, legend columns=2, legend style={draw=none,/tikz/column 2/.style={column sep=5pt,},}, every axis plot/.append style={ultra thick}, legend style = {font=\LARGE,fill=none},ticklabel style={font=\LARGE},title style={font=\LARGE},cycle list name=color, ylabel={winning probability}, xlabel={$\normphi$},every tick label/.append style={font=\Huge}, 
label style={font=\Huge}]
\addplot+[lightgray,forget plot, thick] coordinates{(0,0.5) (0.5,0.5)};
\addplot[red, line width=3pt]coordinates{(0.0, 1.0) (0.0025, 0.9922) (0.005, 0.9723) (0.0075, 0.9437) (0.01, 0.9068) (0.0125, 0.8745) (0.015, 0.8421) (0.0175, 0.7913) (0.02, 0.7637) (0.0225, 0.7294) (0.025, 0.6889) (0.0275, 0.651) (0.03, 0.623) (0.0325, 0.6042) (0.035, 0.5776) (0.0375, 0.5489) (0.04, 0.5299) (0.0425, 0.5013) (0.045, 0.4914) (0.0475, 0.4634) (0.05, 0.4561) (0.0525, 0.4363) (0.055, 0.4207) (0.0575, 0.3955) (0.06, 0.4021) (0.0625, 0.3849) (0.065, 0.3727) (0.0675, 0.3547) (0.07, 0.351) (0.0725, 0.3429) (0.075, 0.3164) (0.0775, 0.3197) (0.08, 0.3131) (0.0825, 0.2986) (0.085, 0.3014) (0.08750000000000001, 0.2865) (0.09, 0.2852) (0.0925, 0.2688) (0.095, 0.2713) (0.0975, 0.2596) (0.1, 0.2626) (0.10250000000000001, 0.2494) (0.105, 0.2365) (0.1075, 0.2416) (0.11, 0.2399) (0.1125, 0.2297) (0.115, 0.2229) (0.11750000000000001, 0.2211) (0.12, 0.2146) (0.1225, 0.2054) (0.125, 0.2019) (0.1275, 0.1993) (0.13, 0.1932) (0.1325, 0.1953) (0.135, 0.1874) (0.1375, 0.1906) (0.14, 0.1836) (0.14250000000000002, 0.188) (0.145, 0.1775) (0.1475, 0.1752) (0.15, 0.1637) (0.1525, 0.1722) (0.155, 0.1737) (0.1575, 0.1633) (0.16, 0.1684) (0.1625, 0.1632) (0.165, 0.1638) (0.1675, 0.1563) (0.17, 0.1578) (0.17250000000000001, 0.1523) (0.17500000000000002, 0.1547) (0.1775, 0.1511) (0.18, 0.1545) (0.1825, 0.1521) (0.185, 0.1483) (0.1875, 0.1472) (0.19, 0.1414) (0.1925, 0.1459) (0.195, 0.1362) (0.1975, 0.1396) (0.2, 0.1451) (0.2025, 0.1391) (0.20500000000000002, 0.1365) (0.20750000000000002, 0.1321) (0.21, 0.1335) (0.2125, 0.1352) (0.215, 0.1272) (0.2175, 0.1268) (0.22, 0.1313) (0.2225, 0.1239) (0.225, 0.1247) (0.2275, 0.122) (0.23, 0.119) (0.2325, 0.1249) (0.23500000000000001, 0.1231) (0.23750000000000002, 0.1236) (0.24, 0.1184) (0.2425, 0.1185) (0.245, 0.1162) (0.2475, 0.1159) (0.25, 0.1072) (0.2525, 0.1187) (0.255, 0.1143) (0.2575, 0.1144) (0.26, 0.1128) (0.2625, 0.1136) (0.265, 0.1143) (0.2675, 0.1103) (0.27, 0.1078) (0.2725, 0.1077) (0.275, 0.1076) (0.2775, 0.1065) (0.28, 0.1053) (0.28250000000000003, 0.1125) (0.28500000000000003, 0.1004) (0.28750000000000003, 0.1056) (0.29, 0.1048) (0.2925, 0.106) (0.295, 0.102) (0.2975, 0.1085) (0.3, 0.1031) (0.3025, 0.1038) (0.305, 0.1038) (0.3075, 0.1045) (0.31, 0.0992) (0.3125, 0.0966) (0.315, 0.1012) (0.3175, 0.1001) (0.32, 0.0951) (0.3225, 0.0992) (0.325, 0.0916) (0.3275, 0.1004) (0.33, 0.1011) (0.3325, 0.0985) (0.335, 0.0993) (0.3375, 0.0979) (0.34, 0.0994) (0.3425, 0.0932) (0.34500000000000003, 0.0957) (0.34750000000000003, 0.0979) (0.35000000000000003, 0.0954) (0.3525, 0.0956) (0.355, 0.0983) (0.3575, 0.0973) (0.36, 0.0937) (0.3625, 0.0956) (0.365, 0.0937) (0.3675, 0.0909) (0.37, 0.0977) (0.3725, 0.097) (0.375, 0.0954) (0.3775, 0.0961) (0.38, 0.0912) (0.3825, 0.0971) (0.385, 0.0916) (0.3875, 0.0951) (0.39, 0.0955) (0.3925, 0.0921) (0.395, 0.0939) (0.3975, 0.0944) (0.4, 0.0938) (0.4025, 0.0922) (0.405, 0.088) (0.40750000000000003, 0.0836) (0.41000000000000003, 0.0965) (0.41250000000000003, 0.0935) (0.41500000000000004, 0.0917) (0.4175, 0.0924) (0.42, 0.0938) (0.4225, 0.0958) (0.425, 0.0912) (0.4275, 0.0928) (0.43, 0.0881) (0.4325, 0.0907) (0.435, 0.0885) (0.4375, 0.093) (0.44, 0.0902) (0.4425, 0.0969) (0.445, 0.0987) (0.4475, 0.0896) (0.45, 0.0901) (0.4525, 0.0879) (0.455, 0.0871) (0.4575, 0.0889) (0.46, 0.0941) (0.4625, 0.0877) (0.465, 0.0913) (0.4675, 0.0932) (0.47000000000000003, 0.088) (0.47250000000000003, 0.0916) (0.47500000000000003, 0.0903) (0.47750000000000004, 0.0936) (0.48, 0.088) (0.4825, 0.0917) (0.485, 0.0893) (0.4875, 0.0918) (0.49, 0.0893) (0.4925, 0.0891) (0.495, 0.0926) (0.4975, 0.0888) (0.5, 0.0941)};
\addplot[blue, line width=3pt]coordinates{(0.0, 0.0) (0.0025, 0.1106) (0.005, 0.1717) (0.0075, 0.2261) (0.01, 0.2536) (0.0125, 0.28) (0.015, 0.2932) (0.0175, 0.3059) (0.02, 0.3084) (0.0225, 0.3074) (0.025, 0.3064) (0.0275, 0.3093) (0.03, 0.299) (0.0325, 0.2946) (0.035, 0.288) (0.0375, 0.29) (0.04, 0.2855) (0.0425, 0.2816) (0.045, 0.2698) (0.0475, 0.2707) (0.05, 0.2613) (0.0525, 0.2593) (0.055, 0.2615) (0.0575, 0.2488) (0.06, 0.2429) (0.0625, 0.2405) (0.065, 0.2412) (0.0675, 0.234) (0.07, 0.2262) (0.0725, 0.2287) (0.075, 0.2246) (0.0775, 0.2198) (0.08, 0.2202) (0.0825, 0.2107) (0.085, 0.2158) (0.08750000000000001, 0.2113) (0.09, 0.2087) (0.0925, 0.2012) (0.095, 0.2013) (0.0975, 0.1989) (0.1, 0.192) (0.10250000000000001, 0.1913) (0.105, 0.1846) (0.1075, 0.1834) (0.11, 0.188) (0.1125, 0.1796) (0.115, 0.1802) (0.11750000000000001, 0.1726) (0.12, 0.1722) (0.1225, 0.1668) (0.125, 0.173) (0.1275, 0.1756) (0.13, 0.1652) (0.1325, 0.1645) (0.135, 0.1641) (0.1375, 0.1591) (0.14, 0.1693) (0.14250000000000002, 0.1573) (0.145, 0.1587) (0.1475, 0.1622) (0.15, 0.1589) (0.1525, 0.146) (0.155, 0.1504) (0.1575, 0.1578) (0.16, 0.1498) (0.1625, 0.1503) (0.165, 0.1443) (0.1675, 0.14) (0.17, 0.1461) (0.17250000000000001, 0.1384) (0.17500000000000002, 0.1413) (0.1775, 0.135) (0.18, 0.1404) (0.1825, 0.1344) (0.185, 0.1368) (0.1875, 0.1381) (0.19, 0.1297) (0.1925, 0.1311) (0.195, 0.1236) (0.1975, 0.1281) (0.2, 0.1258) (0.2025, 0.1252) (0.20500000000000002, 0.1255) (0.20750000000000002, 0.1259) (0.21, 0.1283) (0.2125, 0.1271) (0.215, 0.1228) (0.2175, 0.1197) (0.22, 0.1217) (0.2225, 0.1171) (0.225, 0.1167) (0.2275, 0.1155) (0.23, 0.1112) (0.2325, 0.1143) (0.23500000000000001, 0.1141) (0.23750000000000002, 0.116) (0.24, 0.1147) (0.2425, 0.1127) (0.245, 0.1101) (0.2475, 0.1139) (0.25, 0.1094) (0.2525, 0.1098) (0.255, 0.1098) (0.2575, 0.1064) (0.26, 0.1038) (0.2625, 0.108) (0.265, 0.1045) (0.2675, 0.11) (0.27, 0.1059) (0.2725, 0.1069) (0.275, 0.1008) (0.2775, 0.1033) (0.28, 0.1039) (0.28250000000000003, 0.1015) (0.28500000000000003, 0.1017) (0.28750000000000003, 0.0952) (0.29, 0.1003) (0.2925, 0.1003) (0.295, 0.0993) (0.2975, 0.1042) (0.3, 0.0966) (0.3025, 0.097) (0.305, 0.0927) (0.3075, 0.0969) (0.31, 0.0975) (0.3125, 0.0953) (0.315, 0.0967) (0.3175, 0.0949) (0.32, 0.0892) (0.3225, 0.0917) (0.325, 0.0919) (0.3275, 0.0944) (0.33, 0.0981) (0.3325, 0.0897) (0.335, 0.0915) (0.3375, 0.0886) (0.34, 0.0946) (0.3425, 0.0875) (0.34500000000000003, 0.09) (0.34750000000000003, 0.086) (0.35000000000000003, 0.0929) (0.3525, 0.0883) (0.355, 0.0886) (0.3575, 0.0831) (0.36, 0.0844) (0.3625, 0.0842) (0.365, 0.086) (0.3675, 0.086) (0.37, 0.0859) (0.3725, 0.0876) (0.375, 0.0901) (0.3775, 0.0857) (0.38, 0.0874) (0.3825, 0.084) (0.385, 0.0837) (0.3875, 0.081) (0.39, 0.0826) (0.3925, 0.0814) (0.395, 0.0818) (0.3975, 0.0838) (0.4, 0.0792) (0.4025, 0.0807) (0.405, 0.0769) (0.40750000000000003, 0.077) (0.41000000000000003, 0.0802) (0.41250000000000003, 0.083) (0.41500000000000004, 0.0786) (0.4175, 0.0792) (0.42, 0.0825) (0.4225, 0.0799) (0.425, 0.0784) (0.4275, 0.0833) (0.43, 0.0776) (0.4325, 0.0747) (0.435, 0.0801) (0.4375, 0.0865) (0.44, 0.0799) (0.4425, 0.0785) (0.445, 0.0794) (0.4475, 0.0769) (0.45, 0.0778) (0.4525, 0.076) (0.455, 0.0804) (0.4575, 0.0771) (0.46, 0.0762) (0.4625, 0.0757) (0.465, 0.0813) (0.4675, 0.0778) (0.47000000000000003, 0.0796) (0.47250000000000003, 0.08) (0.47500000000000003, 0.0811) (0.47750000000000004, 0.0806) (0.48, 0.0751) (0.4825, 0.0755) (0.485, 0.0827) (0.4875, 0.0764) (0.49, 0.0742) (0.4925, 0.0753) (0.495, 0.0807) (0.4975, 0.0749) (0.5, 0.0802)};
\addplot[black, line width=3pt]coordinates{(0.0, 0.0) (0.0025, 0.0186) (0.005, 0.0408) (0.0075, 0.0676) (0.01, 0.0888) (0.0125, 0.1071) (0.015, 0.1332) (0.0175, 0.1519) (0.02, 0.1698) (0.0225, 0.1765) (0.025, 0.197) (0.0275, 0.212) (0.03, 0.2116) (0.0325, 0.2232) (0.035, 0.2298) (0.0375, 0.2368) (0.04, 0.2448) (0.0425, 0.2465) (0.045, 0.2584) (0.0475, 0.2593) (0.05, 0.2524) (0.0525, 0.2635) (0.055, 0.2618) (0.0575, 0.276) (0.06, 0.259) (0.0625, 0.2716) (0.065, 0.2702) (0.0675, 0.271) (0.07, 0.2684) (0.0725, 0.2806) (0.075, 0.2777) (0.0775, 0.2764) (0.08, 0.2786) (0.0825, 0.2782) (0.085, 0.2708) (0.08750000000000001, 0.2745) (0.09, 0.2718) (0.0925, 0.2794) (0.095, 0.2727) (0.0975, 0.2797) (0.1, 0.2772) (0.10250000000000001, 0.2751) (0.105, 0.2875) (0.1075, 0.2797) (0.11, 0.2763) (0.1125, 0.2855) (0.115, 0.2731) (0.11750000000000001, 0.2799) (0.12, 0.2768) (0.1225, 0.278) (0.125, 0.2816) (0.1275, 0.2789) (0.13, 0.2759) (0.1325, 0.2802) (0.135, 0.2813) (0.1375, 0.2777) (0.14, 0.2689) (0.14250000000000002, 0.2761) (0.145, 0.2831) (0.1475, 0.2756) (0.15, 0.2788) (0.1525, 0.2679) (0.155, 0.2747) (0.1575, 0.2699) (0.16, 0.2691) (0.1625, 0.2681) (0.165, 0.2629) (0.1675, 0.2686) (0.17, 0.2665) (0.17250000000000001, 0.2633) (0.17500000000000002, 0.269) (0.1775, 0.2635) (0.18, 0.2678) (0.1825, 0.2605) (0.185, 0.2642) (0.1875, 0.2634) (0.19, 0.2648) (0.1925, 0.2632) (0.195, 0.2649) (0.1975, 0.2642) (0.2, 0.2604) (0.2025, 0.2566) (0.20500000000000002, 0.2642) (0.20750000000000002, 0.2661) (0.21, 0.2603) (0.2125, 0.2584) (0.215, 0.2549) (0.2175, 0.2525) (0.22, 0.2543) (0.2225, 0.256) (0.225, 0.2526) (0.2275, 0.2594) (0.23, 0.2505) (0.2325, 0.2461) (0.23500000000000001, 0.2482) (0.23750000000000002, 0.2527) (0.24, 0.2522) (0.2425, 0.2441) (0.245, 0.247) (0.2475, 0.2459) (0.25, 0.2484) (0.2525, 0.2411) (0.255, 0.2396) (0.2575, 0.242) (0.26, 0.2443) (0.2625, 0.2409) (0.265, 0.2491) (0.2675, 0.2385) (0.27, 0.2468) (0.2725, 0.2442) (0.275, 0.2435) (0.2775, 0.2354) (0.28, 0.2388) (0.28250000000000003, 0.2389) (0.28500000000000003, 0.2373) (0.28750000000000003, 0.2365) (0.29, 0.2345) (0.2925, 0.2348) (0.295, 0.2326) (0.2975, 0.2324) (0.3, 0.2339) (0.3025, 0.2287) (0.305, 0.2331) (0.3075, 0.2343) (0.31, 0.2419) (0.3125, 0.2255) (0.315, 0.2239) (0.3175, 0.2262) (0.32, 0.223) (0.3225, 0.2335) (0.325, 0.2336) (0.3275, 0.2303) (0.33, 0.231) (0.3325, 0.2227) (0.335, 0.2205) (0.3375, 0.2244) (0.34, 0.2184) (0.3425, 0.2266) (0.34500000000000003, 0.2253) (0.34750000000000003, 0.218) (0.35000000000000003, 0.2243) (0.3525, 0.2175) (0.355, 0.2308) (0.3575, 0.2218) (0.36, 0.2216) (0.3625, 0.2202) (0.365, 0.2185) (0.3675, 0.2154) (0.37, 0.2171) (0.3725, 0.2105) (0.375, 0.2106) (0.3775, 0.2174) (0.38, 0.2177) (0.3825, 0.2114) (0.385, 0.2101) (0.3875, 0.2155) (0.39, 0.2112) (0.3925, 0.2168) (0.395, 0.2162) (0.3975, 0.224) (0.4, 0.2092) (0.4025, 0.2116) (0.405, 0.2114) (0.40750000000000003, 0.2074) (0.41000000000000003, 0.2102) (0.41250000000000003, 0.2121) (0.41500000000000004, 0.2104) (0.4175, 0.208) (0.42, 0.2015) (0.4225, 0.2045) (0.425, 0.2043) (0.4275, 0.2086) (0.43, 0.2044) (0.4325, 0.2026) (0.435, 0.2069) (0.4375, 0.194) (0.44, 0.2022) (0.4425, 0.1993) (0.445, 0.2021) (0.4475, 0.2061) (0.45, 0.1989) (0.4525, 0.2029) (0.455, 0.1954) (0.4575, 0.2003) (0.46, 0.198) (0.4625, 0.2012) (0.465, 0.1968) (0.4675, 0.1957) (0.47000000000000003, 0.1991) (0.47250000000000003, 0.1969) (0.47500000000000003, 0.1889) (0.47750000000000004, 0.1936) (0.48, 0.1972) (0.4825, 0.1991) (0.485, 0.1898) (0.4875, 0.1931) (0.49, 0.1897) (0.4925, 0.2019) (0.495, 0.1943) (0.4975, 0.195) (0.5, 0.1819)};
\addplot[black!30!green, line width=3pt]coordinates{(0.0, 0.0) (0.0025, 0.001) (0.005, 0.0032) (0.0075, 0.009) (0.01, 0.0139) (0.0125, 0.0251) (0.015, 0.0277) (0.0175, 0.0372) (0.02, 0.0458) (0.0225, 0.0524) (0.025, 0.0641) (0.0275, 0.0659) (0.03, 0.0755) (0.0325, 0.0821) (0.035, 0.0864) (0.0375, 0.095) (0.04, 0.099) (0.0425, 0.1031) (0.045, 0.1044) (0.0475, 0.1127) (0.05, 0.1139) (0.0525, 0.1177) (0.055, 0.1208) (0.0575, 0.124) (0.06, 0.1327) (0.0625, 0.1315) (0.065, 0.1338) (0.0675, 0.1404) (0.07, 0.137) (0.0725, 0.1415) (0.075, 0.1432) (0.0775, 0.1432) (0.08, 0.1484) (0.0825, 0.1494) (0.085, 0.1494) (0.08750000000000001, 0.1588) (0.09, 0.163) (0.0925, 0.1598) (0.095, 0.1609) (0.0975, 0.1667) (0.1, 0.1635) (0.10250000000000001, 0.1734) (0.105, 0.1666) (0.1075, 0.1651) (0.11, 0.1691) (0.1125, 0.1721) (0.115, 0.1781) (0.11750000000000001, 0.1737) (0.12, 0.1828) (0.1225, 0.1766) (0.125, 0.1745) (0.1275, 0.1816) (0.13, 0.1842) (0.1325, 0.1813) (0.135, 0.1852) (0.1375, 0.1807) (0.14, 0.1793) (0.14250000000000002, 0.1799) (0.145, 0.188) (0.1475, 0.1879) (0.15, 0.1912) (0.1525, 0.1938) (0.155, 0.1864) (0.1575, 0.194) (0.16, 0.1956) (0.1625, 0.1907) (0.165, 0.1951) (0.1675, 0.1942) (0.17, 0.1933) (0.17250000000000001, 0.1977) (0.17500000000000002, 0.1929) (0.1775, 0.206) (0.18, 0.1866) (0.1825, 0.1953) (0.185, 0.1984) (0.1875, 0.2007) (0.19, 0.2011) (0.1925, 0.2006) (0.195, 0.2033) (0.1975, 0.211) (0.2, 0.2075) (0.2025, 0.2104) (0.20500000000000002, 0.2071) (0.20750000000000002, 0.2067) (0.21, 0.2035) (0.2125, 0.2042) (0.215, 0.2086) (0.2175, 0.2075) (0.22, 0.2066) (0.2225, 0.2082) (0.225, 0.2074) (0.2275, 0.2028) (0.23, 0.2154) (0.2325, 0.2113) (0.23500000000000001, 0.2067) (0.23750000000000002, 0.2141) (0.24, 0.2116) (0.2425, 0.2172) (0.245, 0.2121) (0.2475, 0.2072) (0.25, 0.2071) (0.2525, 0.2176) (0.255, 0.2106) (0.2575, 0.2199) (0.26, 0.2137) (0.2625, 0.2139) (0.265, 0.2111) (0.2675, 0.2111) (0.27, 0.2089) (0.2725, 0.2117) (0.275, 0.2161) (0.2775, 0.2094) (0.28, 0.2176) (0.28250000000000003, 0.2106) (0.28500000000000003, 0.2119) (0.28750000000000003, 0.2141) (0.29, 0.2236) (0.2925, 0.2203) (0.295, 0.2248) (0.2975, 0.2215) (0.3, 0.2213) (0.3025, 0.2198) (0.305, 0.2225) (0.3075, 0.2228) (0.31, 0.2258) (0.3125, 0.2296) (0.315, 0.2236) (0.3175, 0.2221) (0.32, 0.22) (0.3225, 0.2171) (0.325, 0.2213) (0.3275, 0.2192) (0.33, 0.2163) (0.3325, 0.2194) (0.335, 0.2242) (0.3375, 0.2176) (0.34, 0.2161) (0.3425, 0.2256) (0.34500000000000003, 0.2222) (0.34750000000000003, 0.2325) (0.35000000000000003, 0.2245) (0.3525, 0.2205) (0.355, 0.228) (0.3575, 0.2267) (0.36, 0.2266) (0.3625, 0.2256) (0.365, 0.2264) (0.3675, 0.2232) (0.37, 0.2349) (0.3725, 0.2236) (0.375, 0.2353) (0.3775, 0.2251) (0.38, 0.2243) (0.3825, 0.2245) (0.385, 0.2253) (0.3875, 0.2283) (0.39, 0.2303) (0.3925, 0.2312) (0.395, 0.2246) (0.3975, 0.2273) (0.4, 0.2214) (0.4025, 0.2289) (0.405, 0.232) (0.40750000000000003, 0.2294) (0.41000000000000003, 0.2304) (0.41250000000000003, 0.2269) (0.41500000000000004, 0.2154) (0.4175, 0.2257) (0.42, 0.2367) (0.4225, 0.2342) (0.425, 0.2289) (0.4275, 0.2322) (0.43, 0.2297) (0.4325, 0.2272) (0.435, 0.2339) (0.4375, 0.2249) (0.44, 0.2319) (0.4425, 0.2276) (0.445, 0.2266) (0.4475, 0.2272) (0.45, 0.23) (0.4525, 0.2292) (0.455, 0.2262) (0.4575, 0.2284) (0.46, 0.231) (0.4625, 0.2231) (0.465, 0.2285) (0.4675, 0.2302) (0.47000000000000003, 0.2241) (0.47250000000000003, 0.2254) (0.47500000000000003, 0.231) (0.47750000000000004, 0.2269) (0.48, 0.2324) (0.4825, 0.2265) (0.485, 0.232) (0.4875, 0.2242) (0.49, 0.2274) (0.4925, 0.2279) (0.495, 0.2252) (0.4975, 0.229) (0.5, 0.2285)};
\end{axis}

				\end{tikzpicture}}
			\caption{Election generated from Mallows model with lexicographic central order and  $\normphi=1$}
			\label{fig:IC}
		\end{subfigure}\hfil 
		\begin{subfigure}{0.45\textwidth}
			\centering 
			\resizebox{0.9\textwidth}{!}{\begin{tikzpicture}
\begin{axis}[anchor=north, no markers, enlargelimits=false, legend columns=2, legend style={draw=none,/tikz/column 2/.style={column sep=5pt,},}, every axis plot/.append style={ultra thick}, legend style = {font=\LARGE,fill=none},ticklabel style={font=\LARGE},title style={font=\LARGE},cycle list name=color list, ylabel={winning probability}, xlabel={$\normphi$},every tick label/.append style={font=\Huge}, 
label style={font=\Huge}]
\addplot+[lightgray,forget plot, thick] coordinates{(0,0.5) (0.5,0.5)};
\addplot[red, line width=3pt]coordinates{(0.0, 1.0) (0.0025, 1.0) (0.005, 0.9996) (0.0075, 0.9998) (0.01, 0.9997) (0.0125, 0.9986) (0.015, 0.9982) (0.0175, 0.9971) (0.02, 0.9947) (0.0225, 0.9939) (0.025, 0.9906) (0.0275, 0.989) (0.03, 0.9852) (0.0325, 0.9819) (0.035, 0.9787) (0.0375, 0.9762) (0.04, 0.9726) (0.0425, 0.9698) (0.045, 0.9622) (0.0475, 0.9599) (0.05, 0.9522) (0.0525, 0.9456) (0.055, 0.9429) (0.0575, 0.9357) (0.06, 0.9325) (0.0625, 0.9316) (0.065, 0.9335) (0.0675, 0.9222) (0.07, 0.9209) (0.0725, 0.9149) (0.075, 0.9163) (0.0775, 0.9064) (0.08, 0.9051) (0.0825, 0.9043) (0.085, 0.8929) (0.08750000000000001, 0.8902) (0.09, 0.8841) (0.0925, 0.8886) (0.095, 0.8846) (0.0975, 0.8741) (0.1, 0.8726) (0.10250000000000001, 0.8736) (0.105, 0.8653) (0.1075, 0.8625) (0.11, 0.8632) (0.1125, 0.8471) (0.115, 0.8456) (0.11750000000000001, 0.848) (0.12, 0.8435) (0.1225, 0.8442) (0.125, 0.8417) (0.1275, 0.8325) (0.13, 0.8395) (0.1325, 0.8272) (0.135, 0.8228) (0.1375, 0.8265) (0.14, 0.8192) (0.14250000000000002, 0.8187) (0.145, 0.8139) (0.1475, 0.8153) (0.15, 0.8061) (0.1525, 0.8096) (0.155, 0.8067) (0.1575, 0.8033) (0.16, 0.7971) (0.1625, 0.8055) (0.165, 0.791) (0.1675, 0.7904) (0.17, 0.778) (0.17250000000000001, 0.7869) (0.17500000000000002, 0.7911) (0.1775, 0.7851) (0.18, 0.7756) (0.1825, 0.7728) (0.185, 0.7764) (0.1875, 0.772) (0.19, 0.7748) (0.1925, 0.7669) (0.195, 0.765) (0.1975, 0.7613) (0.2, 0.7517) (0.2025, 0.7529) (0.20500000000000002, 0.7616) (0.20750000000000002, 0.7567) (0.21, 0.7492) (0.2125, 0.7589) (0.215, 0.7425) (0.2175, 0.7393) (0.22, 0.7407) (0.2225, 0.7372) (0.225, 0.7399) (0.2275, 0.7329) (0.23, 0.7348) (0.2325, 0.7261) (0.23500000000000001, 0.7273) (0.23750000000000002, 0.7233) (0.24, 0.7253) (0.2425, 0.7212) (0.245, 0.7226) (0.2475, 0.7153) (0.25, 0.7084) (0.2525, 0.7148) (0.255, 0.7058) (0.2575, 0.7152) (0.26, 0.7105) (0.2625, 0.7063) (0.265, 0.7093) (0.2675, 0.7075) (0.27, 0.7045) (0.2725, 0.7058) (0.275, 0.7013) (0.2775, 0.7012) (0.28, 0.6891) (0.28250000000000003, 0.6894) (0.28500000000000003, 0.6843) (0.28750000000000003, 0.6918) (0.29, 0.6891) (0.2925, 0.6924) (0.295, 0.6815) (0.2975, 0.6722) (0.3, 0.6803) (0.3025, 0.6827) (0.305, 0.6765) (0.3075, 0.6699) (0.31, 0.6678) (0.3125, 0.6674) (0.315, 0.6692) (0.3175, 0.6683) (0.32, 0.669) (0.3225, 0.6577) (0.325, 0.6559) (0.3275, 0.6539) (0.33, 0.6568) (0.3325, 0.6553) (0.335, 0.6499) (0.3375, 0.654) (0.34, 0.6474) (0.3425, 0.6558) (0.34500000000000003, 0.6658) (0.34750000000000003, 0.6444) (0.35000000000000003, 0.6451) (0.3525, 0.6364) (0.355, 0.6369) (0.3575, 0.638) (0.36, 0.635) (0.3625, 0.6369) (0.365, 0.6291) (0.3675, 0.6256) (0.37, 0.6258) (0.3725, 0.6352) (0.375, 0.6374) (0.3775, 0.6241) (0.38, 0.6211) (0.3825, 0.6305) (0.385, 0.619) (0.3875, 0.6208) (0.39, 0.6178) (0.3925, 0.6123) (0.395, 0.6164) (0.3975, 0.6104) (0.4, 0.6163) (0.4025, 0.6182) (0.405, 0.6106) (0.40750000000000003, 0.611) (0.41000000000000003, 0.6097) (0.41250000000000003, 0.6018) (0.41500000000000004, 0.6074) (0.4175, 0.5926) (0.42, 0.598) (0.4225, 0.6055) (0.425, 0.6018) (0.4275, 0.5996) (0.43, 0.5942) (0.4325, 0.5885) (0.435, 0.5834) (0.4375, 0.5897) (0.44, 0.5845) (0.4425, 0.5878) (0.445, 0.5854) (0.4475, 0.5767) (0.45, 0.5892) (0.4525, 0.577) (0.455, 0.5711) (0.4575, 0.5757) (0.46, 0.57) (0.4625, 0.5711) (0.465, 0.577) (0.4675, 0.5677) (0.47000000000000003, 0.5659) (0.47250000000000003, 0.5695) (0.47500000000000003, 0.5623) (0.47750000000000004, 0.5689) (0.48, 0.5603) (0.4825, 0.557) (0.485, 0.5502) (0.4875, 0.5462) (0.49, 0.5543) (0.4925, 0.5548) (0.495, 0.5569) (0.4975, 0.5502) (0.5, 0.55)};
\addplot[blue, line width=3pt]coordinates{(0.0, 0.0) (0.0025, 0.0) (0.005, 0.0002) (0.0075, 0.0004) (0.01, 0.0007) (0.0125, 0.0023) (0.015, 0.0039) (0.0175, 0.0039) (0.02, 0.0072) (0.0225, 0.0081) (0.025, 0.0146) (0.0275, 0.0139) (0.03, 0.0195) (0.0325, 0.0221) (0.035, 0.0256) (0.0375, 0.0274) (0.04, 0.0323) (0.0425, 0.0311) (0.045, 0.0394) (0.0475, 0.0431) (0.05, 0.045) (0.0525, 0.0515) (0.055, 0.0521) (0.0575, 0.0575) (0.06, 0.0593) (0.0625, 0.0614) (0.065, 0.0595) (0.0675, 0.0717) (0.07, 0.0664) (0.0725, 0.0744) (0.075, 0.0712) (0.0775, 0.0813) (0.08, 0.0781) (0.0825, 0.0784) (0.085, 0.0859) (0.08750000000000001, 0.0913) (0.09, 0.0921) (0.0925, 0.0917) (0.095, 0.0884) (0.0975, 0.0974) (0.1, 0.1033) (0.10250000000000001, 0.1024) (0.105, 0.1036) (0.1075, 0.1062) (0.11, 0.1056) (0.1125, 0.1084) (0.115, 0.1179) (0.11750000000000001, 0.11) (0.12, 0.1172) (0.1225, 0.1157) (0.125, 0.1195) (0.1275, 0.1204) (0.13, 0.1212) (0.1325, 0.1256) (0.135, 0.1324) (0.1375, 0.1233) (0.14, 0.1317) (0.14250000000000002, 0.1308) (0.145, 0.137) (0.1475, 0.1347) (0.15, 0.1402) (0.1525, 0.1355) (0.155, 0.1411) (0.1575, 0.1358) (0.16, 0.1456) (0.1625, 0.1373) (0.165, 0.1415) (0.1675, 0.1493) (0.17, 0.1533) (0.17250000000000001, 0.1512) (0.17500000000000002, 0.1492) (0.1775, 0.1538) (0.18, 0.1526) (0.1825, 0.1507) (0.185, 0.156) (0.1875, 0.155) (0.19, 0.1526) (0.1925, 0.1611) (0.195, 0.1635) (0.1975, 0.166) (0.2, 0.1739) (0.2025, 0.1642) (0.20500000000000002, 0.1632) (0.20750000000000002, 0.1653) (0.21, 0.1723) (0.2125, 0.1626) (0.215, 0.1746) (0.2175, 0.1756) (0.22, 0.1755) (0.2225, 0.1778) (0.225, 0.177) (0.2275, 0.1768) (0.23, 0.1826) (0.2325, 0.1829) (0.23500000000000001, 0.1825) (0.23750000000000002, 0.1822) (0.24, 0.1834) (0.2425, 0.1823) (0.245, 0.1837) (0.2475, 0.1902) (0.25, 0.1922) (0.2525, 0.189) (0.255, 0.1943) (0.2575, 0.1929) (0.26, 0.1873) (0.2625, 0.1965) (0.265, 0.1927) (0.2675, 0.1966) (0.27, 0.1972) (0.2725, 0.1901) (0.275, 0.1925) (0.2775, 0.1952) (0.28, 0.2019) (0.28250000000000003, 0.1986) (0.28500000000000003, 0.2025) (0.28750000000000003, 0.1997) (0.29, 0.199) (0.2925, 0.2005) (0.295, 0.2025) (0.2975, 0.2136) (0.3, 0.2041) (0.3025, 0.2008) (0.305, 0.2053) (0.3075, 0.2084) (0.31, 0.216) (0.3125, 0.2099) (0.315, 0.2104) (0.3175, 0.2145) (0.32, 0.2125) (0.3225, 0.2183) (0.325, 0.2141) (0.3275, 0.2189) (0.33, 0.217) (0.3325, 0.2139) (0.335, 0.2176) (0.3375, 0.2158) (0.34, 0.2243) (0.3425, 0.2158) (0.34500000000000003, 0.2163) (0.34750000000000003, 0.2209) (0.35000000000000003, 0.2251) (0.3525, 0.2309) (0.355, 0.2194) (0.3575, 0.2283) (0.36, 0.2309) (0.3625, 0.2247) (0.365, 0.2246) (0.3675, 0.2302) (0.37, 0.2298) (0.3725, 0.2233) (0.375, 0.221) (0.3775, 0.2312) (0.38, 0.231) (0.3825, 0.2305) (0.385, 0.2298) (0.3875, 0.2343) (0.39, 0.2373) (0.3925, 0.2319) (0.395, 0.242) (0.3975, 0.2279) (0.4, 0.2359) (0.4025, 0.2343) (0.405, 0.2395) (0.40750000000000003, 0.236) (0.41000000000000003, 0.236) (0.41250000000000003, 0.2395) (0.41500000000000004, 0.2382) (0.4175, 0.2436) (0.42, 0.246) (0.4225, 0.2417) (0.425, 0.229) (0.4275, 0.2432) (0.43, 0.2416) (0.4325, 0.248) (0.435, 0.2397) (0.4375, 0.2446) (0.44, 0.2456) (0.4425, 0.2439) (0.445, 0.2458) (0.4475, 0.2468) (0.45, 0.2383) (0.4525, 0.248) (0.455, 0.2519) (0.4575, 0.2482) (0.46, 0.2493) (0.4625, 0.2452) (0.465, 0.2469) (0.4675, 0.253) (0.47000000000000003, 0.2486) (0.47250000000000003, 0.2497) (0.47500000000000003, 0.2539) (0.47750000000000004, 0.2448) (0.48, 0.2548) (0.4825, 0.2553) (0.485, 0.2569) (0.4875, 0.2548) (0.49, 0.2518) (0.4925, 0.2492) (0.495, 0.258) (0.4975, 0.249) (0.5, 0.2526)};
\addplot[black, line width=3pt]coordinates{(0.0, 0.0) (0.0025, 0.0) (0.005, 0.0006) (0.0075, 0.0005) (0.01, 0.001) (0.0125, 0.0021) (0.015, 0.0021) (0.0175, 0.0042) (0.02, 0.0065) (0.0225, 0.0065) (0.025, 0.0088) (0.0275, 0.0112) (0.03, 0.0126) (0.0325, 0.0186) (0.035, 0.0189) (0.0375, 0.0194) (0.04, 0.0228) (0.0425, 0.0268) (0.045, 0.0291) (0.0475, 0.0279) (0.05, 0.0353) (0.0525, 0.0351) (0.055, 0.036) (0.0575, 0.039) (0.06, 0.0428) (0.0625, 0.0422) (0.065, 0.047) (0.0675, 0.0493) (0.07, 0.0483) (0.0725, 0.0506) (0.075, 0.0534) (0.0775, 0.0543) (0.08, 0.0573) (0.0825, 0.0632) (0.085, 0.0615) (0.08750000000000001, 0.0618) (0.09, 0.0657) (0.0925, 0.063) (0.095, 0.0672) (0.0975, 0.0709) (0.1, 0.0696) (0.10250000000000001, 0.0737) (0.105, 0.0717) (0.1075, 0.079) (0.11, 0.0743) (0.1125, 0.0848) (0.115, 0.084) (0.11750000000000001, 0.0864) (0.12, 0.0816) (0.1225, 0.0851) (0.125, 0.0793) (0.1275, 0.0852) (0.13, 0.0843) (0.1325, 0.0922) (0.135, 0.0877) (0.1375, 0.096) (0.14, 0.0932) (0.14250000000000002, 0.095) (0.145, 0.0947) (0.1475, 0.0939) (0.15, 0.0969) (0.1525, 0.0986) (0.155, 0.0986) (0.1575, 0.1018) (0.16, 0.1046) (0.1625, 0.0982) (0.165, 0.104) (0.1675, 0.0985) (0.17, 0.1098) (0.17250000000000001, 0.1076) (0.17500000000000002, 0.105) (0.1775, 0.1076) (0.18, 0.112) (0.1825, 0.1149) (0.185, 0.1108) (0.1875, 0.121) (0.19, 0.1097) (0.1925, 0.1169) (0.195, 0.1095) (0.1975, 0.1161) (0.2, 0.1187) (0.2025, 0.1211) (0.20500000000000002, 0.1186) (0.20750000000000002, 0.1167) (0.21, 0.1233) (0.2125, 0.1225) (0.215, 0.1244) (0.2175, 0.1286) (0.22, 0.1185) (0.2225, 0.1304) (0.225, 0.1281) (0.2275, 0.1306) (0.23, 0.1248) (0.2325, 0.1314) (0.23500000000000001, 0.1312) (0.23750000000000002, 0.1318) (0.24, 0.1312) (0.2425, 0.1348) (0.245, 0.1388) (0.2475, 0.1322) (0.25, 0.1402) (0.2525, 0.1362) (0.255, 0.1451) (0.2575, 0.1363) (0.26, 0.1407) (0.2625, 0.1372) (0.265, 0.138) (0.2675, 0.1357) (0.27, 0.14) (0.2725, 0.1401) (0.275, 0.1368) (0.2775, 0.1436) (0.28, 0.1413) (0.28250000000000003, 0.1454) (0.28500000000000003, 0.1497) (0.28750000000000003, 0.1473) (0.29, 0.148) (0.2925, 0.1452) (0.295, 0.1474) (0.2975, 0.1475) (0.3, 0.1489) (0.3025, 0.1522) (0.305, 0.1552) (0.3075, 0.15) (0.31, 0.1506) (0.3125, 0.1636) (0.315, 0.1544) (0.3175, 0.1515) (0.32, 0.1559) (0.3225, 0.1576) (0.325, 0.1584) (0.3275, 0.1535) (0.33, 0.1583) (0.3325, 0.162) (0.335, 0.1624) (0.3375, 0.1575) (0.34, 0.1613) (0.3425, 0.16) (0.34500000000000003, 0.1529) (0.34750000000000003, 0.1672) (0.35000000000000003, 0.1663) (0.3525, 0.1628) (0.355, 0.1712) (0.3575, 0.1665) (0.36, 0.1668) (0.3625, 0.1666) (0.365, 0.1733) (0.3675, 0.1717) (0.37, 0.171) (0.3725, 0.1587) (0.375, 0.1674) (0.3775, 0.1706) (0.38, 0.1752) (0.3825, 0.1772) (0.385, 0.1744) (0.3875, 0.1754) (0.39, 0.1729) (0.3925, 0.177) (0.395, 0.1783) (0.3975, 0.1846) (0.4, 0.1766) (0.4025, 0.1781) (0.405, 0.1734) (0.40750000000000003, 0.1704) (0.41000000000000003, 0.1846) (0.41250000000000003, 0.1846) (0.41500000000000004, 0.1761) (0.4175, 0.1818) (0.42, 0.1791) (0.4225, 0.1802) (0.425, 0.1858) (0.4275, 0.1902) (0.43, 0.1843) (0.4325, 0.1811) (0.435, 0.1903) (0.4375, 0.1813) (0.44, 0.1902) (0.4425, 0.1935) (0.445, 0.1885) (0.4475, 0.188) (0.45, 0.1874) (0.4525, 0.1877) (0.455, 0.1895) (0.4575, 0.1923) (0.46, 0.197) (0.4625, 0.1941) (0.465, 0.1939) (0.4675, 0.1948) (0.47000000000000003, 0.1973) (0.47250000000000003, 0.1927) (0.47500000000000003, 0.1928) (0.47750000000000004, 0.1925) (0.48, 0.1982) (0.4825, 0.1997) (0.485, 0.2004) (0.4875, 0.2054) (0.49, 0.2023) (0.4925, 0.2039) (0.495, 0.1934) (0.4975, 0.2031) (0.5, 0.2024)};
\addplot[black!30!green, line width=3pt]coordinates{(0.0, 0.0) (0.0025, 0.0) (0.005, 0.0) (0.0075, 0.0001) (0.01, 0.0001) (0.0125, 0.0011) (0.015, 0.0011) (0.0175, 0.0028) (0.02, 0.0025) (0.0225, 0.0042) (0.025, 0.0044) (0.0275, 0.0053) (0.03, 0.0057) (0.0325, 0.0084) (0.035, 0.0103) (0.0375, 0.0112) (0.04, 0.0111) (0.0425, 0.0119) (0.045, 0.0154) (0.0475, 0.0157) (0.05, 0.0184) (0.0525, 0.0189) (0.055, 0.0189) (0.0575, 0.022) (0.06, 0.0229) (0.0625, 0.0215) (0.065, 0.0217) (0.0675, 0.0242) (0.07, 0.024) (0.0725, 0.0276) (0.075, 0.0278) (0.0775, 0.0293) (0.08, 0.0292) (0.0825, 0.0272) (0.085, 0.0305) (0.08750000000000001, 0.0331) (0.09, 0.0341) (0.0925, 0.031) (0.095, 0.0337) (0.0975, 0.0367) (0.1, 0.0343) (0.10250000000000001, 0.0356) (0.105, 0.0387) (0.1075, 0.0338) (0.11, 0.0379) (0.1125, 0.0409) (0.115, 0.0421) (0.11750000000000001, 0.043) (0.12, 0.0423) (0.1225, 0.039) (0.125, 0.0427) (0.1275, 0.0466) (0.13, 0.043) (0.1325, 0.0449) (0.135, 0.0438) (0.1375, 0.0418) (0.14, 0.0421) (0.14250000000000002, 0.0444) (0.145, 0.0473) (0.1475, 0.0474) (0.15, 0.0474) (0.1525, 0.046) (0.155, 0.0483) (0.1575, 0.0476) (0.16, 0.0496) (0.1625, 0.0521) (0.165, 0.049) (0.1675, 0.0548) (0.17, 0.0529) (0.17250000000000001, 0.0491) (0.17500000000000002, 0.0483) (0.1775, 0.048) (0.18, 0.0534) (0.1825, 0.055) (0.185, 0.0532) (0.1875, 0.0517) (0.19, 0.0507) (0.1925, 0.0487) (0.195, 0.0552) (0.1975, 0.0522) (0.2, 0.0534) (0.2025, 0.0539) (0.20500000000000002, 0.0564) (0.20750000000000002, 0.0527) (0.21, 0.056) (0.2125, 0.0575) (0.215, 0.0583) (0.2175, 0.056) (0.22, 0.0589) (0.2225, 0.0561) (0.225, 0.0587) (0.2275, 0.0579) (0.23, 0.061) (0.2325, 0.057) (0.23500000000000001, 0.0624) (0.23750000000000002, 0.0652) (0.24, 0.0573) (0.2425, 0.0603) (0.245, 0.0637) (0.2475, 0.0665) (0.25, 0.0635) (0.2525, 0.0632) (0.255, 0.063) (0.2575, 0.0635) (0.26, 0.0618) (0.2625, 0.0634) (0.265, 0.0666) (0.2675, 0.0674) (0.27, 0.0658) (0.2725, 0.0616) (0.275, 0.068) (0.2775, 0.062) (0.28, 0.0695) (0.28250000000000003, 0.0684) (0.28500000000000003, 0.0695) (0.28750000000000003, 0.0669) (0.29, 0.069) (0.2925, 0.0745) (0.295, 0.0693) (0.2975, 0.0675) (0.3, 0.0706) (0.3025, 0.0716) (0.305, 0.072) (0.3075, 0.0758) (0.31, 0.0738) (0.3125, 0.0715) (0.315, 0.0717) (0.3175, 0.0704) (0.32, 0.0716) (0.3225, 0.0724) (0.325, 0.0773) (0.3275, 0.0744) (0.33, 0.074) (0.3325, 0.0776) (0.335, 0.0765) (0.3375, 0.0765) (0.34, 0.0766) (0.3425, 0.0774) (0.34500000000000003, 0.0727) (0.34750000000000003, 0.0741) (0.35000000000000003, 0.0801) (0.3525, 0.082) (0.355, 0.083) (0.3575, 0.0769) (0.36, 0.0794) (0.3625, 0.0796) (0.365, 0.0861) (0.3675, 0.0831) (0.37, 0.0804) (0.3725, 0.0792) (0.375, 0.0812) (0.3775, 0.0841) (0.38, 0.0808) (0.3825, 0.0761) (0.385, 0.0861) (0.3875, 0.086) (0.39, 0.0851) (0.3925, 0.083) (0.395, 0.0796) (0.3975, 0.0933) (0.4, 0.0857) (0.4025, 0.0814) (0.405, 0.088) (0.40750000000000003, 0.0881) (0.41000000000000003, 0.0855) (0.41250000000000003, 0.0862) (0.41500000000000004, 0.0876) (0.4175, 0.088) (0.42, 0.0923) (0.4225, 0.0856) (0.425, 0.0908) (0.4275, 0.0886) (0.43, 0.0899) (0.4325, 0.0916) (0.435, 0.0964) (0.4375, 0.0973) (0.44, 0.0947) (0.4425, 0.0927) (0.445, 0.092) (0.4475, 0.0975) (0.45, 0.0951) (0.4525, 0.0974) (0.455, 0.101) (0.4575, 0.0978) (0.46, 0.0977) (0.4625, 0.1038) (0.465, 0.1027) (0.4675, 0.1016) (0.47000000000000003, 0.0998) (0.47250000000000003, 0.0997) (0.47500000000000003, 0.1055) (0.47750000000000004, 0.1042) (0.48, 0.1058) (0.4825, 0.1057) (0.485, 0.1016) (0.4875, 0.102) (0.49, 0.1053) (0.4925, 0.1077) (0.495, 0.1052) (0.4975, 0.1073) (0.5, 0.107)};
max\end{axis}
  
				\end{tikzpicture}}
			\caption{Election generated from Mallows model with lexicographic central order and $\normphi=0.6$}
			\label{fig:MAL}
		\end{subfigure}     
		\caption{We plot $P_{E,c}(\normphi)$ ($y$-axis) for the Plurality voting rule as a function of $\normphi$ ($x$-axis) for the
			four most successful candidates.} 
		\label{fig:example}
	\end{minipage}\hfill
	\begin{minipage}[b]{0.38\textwidth}
		\resizebox{0.8\textwidth}{!}{\includegraphics[width=\textwidth]{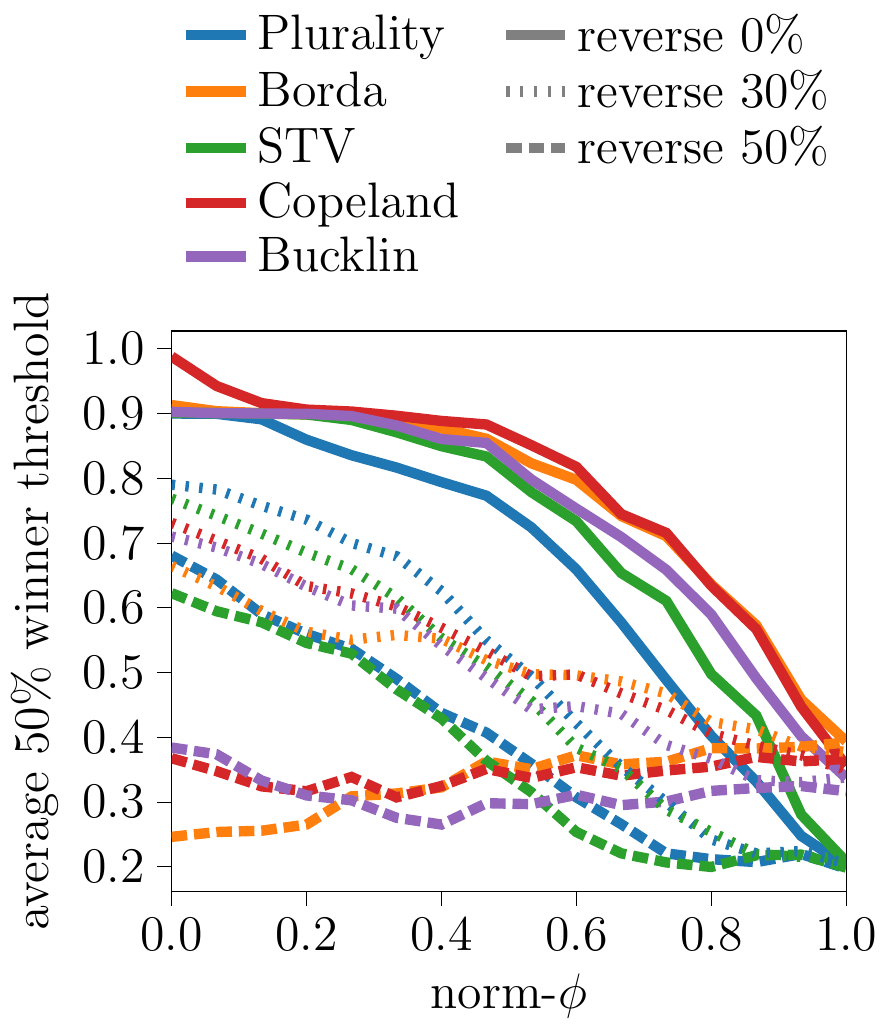}}
		\caption{Average $50\%$-winner threshold of different voting rules for elections sampled from the Mallows model with varying $\normphi$ where each sampled vote is reversed with some probability.}
		\label{fig:robustrules} \label{fig:diverse}
	\end{minipage}
\end{figure}

Our approach relies on the Mallows model, which is typically considered as a natural way to add random noise to an election. 
To model this, we replace each vote $v$ from the election with a vote sampled from the Mallows distribution with central vote $v$ and some normalized dispersion parameter $\normphi\in [0,1]$. 
Specifically, for an election $E=(C,V)$, a candidate $c\in C$, and $\normphi\in [0,1]$, we let $P_{E,c}(\normphi)$ be the probability that candidate $c$ is a winner of an election that results from replacing each vote $v\in V$ with a vote sampled from $\mathcal{D}_{\text{Mallows}}^{v,\normphi}$.
We refer to $P_{E,c}(\normphi)$ as $c$'s winning probability at $\normphi$ and to $1-P_{E,c}(\normphi)$ as $c$'s loosing probability, i.e., the probability that $c$ is not a winner. 
Notably, for $\normphi=1$ each vote has the same probability under the Mallows distribution and, thus, each election has the same probability of being sampled. 
This implies that, assuming votes are complete, all candidates have the same probability of being a winner at $\normphi=1$.
In the following, we say that a winner is robust if $P_{E,c}(\normphi)$ does not ``quickly'' decrease. 
We often visualize the winning probabilities of different candidates as line plots. 
In those plots, each line represents one candidate and depicts its winning probability $P_{E,c}(\normphi)$ (y-axis) for different values of $\normphi$ (x-axis). 
We only depict the range $\normphi\in [0,0.5]$ as for larger values of $\normphi$ the sampled elections have less and less similarities to the given one.
We depict two example plots in \Cref{fig:example}. 
The election displayed in \Cref{fig:IC} is sampled from the Mallows model with $\normphi=1$ (so each vote had the same probability of being sampled). 
This is also clearly visible in the plot: The winning probability of the initially winning red candidate quickly decreases. 
In \Cref{fig:MAL}, we show a more structured election (sampled from the Mallows model with $\normphi=0.6$), where the winning probability of the initially winning red candidate stays high even if substantial random noise is introduced.

Comparing our approach to previous works, \citet{DBLP:conf/ijcai/BoehmerBFN21} followed a related path by computing for a given election $E=(C,V)$ and candidate $c\in C$ the probability $Q_{E,c}(r)$ that $c$ is a winner of an election at swap distance~$r$ from $E$. 
$P_{E,c}(\normphi)$ and $Q_{E,c}(r)$ are indeed closely related because $P_{E,c}(\normphi)$ is a weighted average over $Q_{E,c}(r)$ for different values of $r$, as shown by \citet{DBLP:conf/eumas/BaumeisterH21s}. 
From a computational perspective, computing $Q_{E,c}(r)$ (and thus $P_{E,c}(\normphi)$) exactly is equivalent to solving an instance of $\#\textsc{Swap-Bribery}$ (simply take the number of elections at swap distance $r$ from $E$ where $c$ wins and divide it by the total number of elections at swap distance~$r$). 

Unfortunately, from the results of \citet{DBLP:conf/ijcai/BoehmerBFN21} and \citet{DBLP:conf/eumas/BaumeisterH21s} it follows that solving $\#\textsc{Swap-Bribery}$ and, thus, computing $P_{E,c}(\normphi)$ is intractable. 
This is why we resort to a sampling approach:
To compute $P_{E,c}(\normphi)$ for some $E=(C,V)$,  we sample an election by replacing each vote $v\in V$ by a vote sampled from  $\mathcal{D}_{\text{Mallows}}^{v,\normphi}$. 
We repeat this multiple times and record for each candidate the fraction of sampled elections in which $c$ is a winner.\footnote{By default, for each election we computed $P_{E,c}(\normphi)$ for $\normphi\in \{0,0.1,\dots, 1\}$. 
For each value of $\normphi$, we did so by sampling $500$ elections and recording for each candidate the fraction of these elections where it is a winner.
To evaluate whether $500$ elections are sufficient here, we also reran some of our experiments with $4000$ elections sampled for each value of $\normphi$ and found that the results only marginally changed (in particular, in all elections, the $50\%$-winner threshold changed by at most $0.1$, which is the smallest observable change).
For all visualized elections, we used a finer resolution by computing $P_{E,c}(\normphi)$ for $\normphi=0.0025\cdot i$ for $i\in \{0,1,2,\dots, 200\}$ by sampling for each value of $\normphi$ $\num{10000}$ elections. 
}
To quantify the robustness of a non-tied election $E$, we use the $50\%$-winner threshold introduced by \citet{DBLP:conf/ijcai/BoehmerBFN21}, which is the smallest value of $\normphi$ such that the winning probability of the winner of $E$ is smaller than $50\%$.\footnote{For STV, we cannot simply compute $P_{E,c}(\normphi)$ by sampling some elections and recording in how many of them $c$ is a winner, because deciding whether some candidate is a winner under STV in some given election is NP-hard \cite{DBLP:conf/ijcai/ConitzerRX09}.
Thus, a tie-breaking rule needs to be specified. 
To deal with this issue, for each run of STV on some election, we sample a strict total order $\succ_t$ over all candidates uniformly at random from the set of all strict total orders and break ties according to $\succ_t$.
This in particular implies that as we do $500$ runs at $\normphi=0$, i.e., we apply STV $500$ times to the initial election with different tie-breaking orders, multiple candidates may have a non-zero winning probability in the initial election. 
We consider as the initial winner the candidate having the highest winning probability at $\normphi=0$ and for elections where no candidate has a winning probability over $50\%$ at $\normphi=0$, we set the $50\%$-winner threshold to $0$. \label{stv:footnote}}
The $50\%$-winner threshold thus quantifies how fast the winning probability of the initial winner declines when we move further and further away from the initial election and can be easily used to compare the robustness of election winners in different elections.
Of course, instead of considering the $50\%$-winner threshold, one could also consider  the $x\%$-winner threshold (i.e., the smallest value of $\normphi$ such that the winning probability $P_{E,c}(\normphi)$ of the winner $c$ of $E$ is smaller than $x\%$) for other values of $x$. 
However, for all considered voting rules, the $50\%$-winner threshold is strongly correlated with the $25\%$-winner and $75\%$-winner threshold on the diverse synthetic dataset of \citet{DBLP:conf/atal/SzufaFSST20} (the PCC is typically between $0.85$ and $0.95$). 
As fixing a single value is advantageous for clarity, we picked the $50\%$-winner threshold, since it has a special appeal as it quantifies the perturbation level until which the initial winner is stronger than all other candidates combined. 

The main reason why we use $P_{E,c}(\normphi)$ instead of $Q_{E,c}(r)$, as done by \citet{DBLP:conf/ijcai/BoehmerBFN21}, is that to compute $Q_{E,c}(r)$ we need to sample elections that are exactly at a given swap distance from $E$. 
Unfortunately, this sampling procedure is non-trivial and for more than $20$ candidates already takes quite some time to compute \cite{DBLP:conf/ijcai/BoehmerBFN21}. 
In contrast to this, the approach used in this paper is much faster.
Furthermore, as already argued above, both $P_{E,c}(\normphi)$ and $Q_{E,c}(r)$ are conceptually closely related. 
In particular, each value of $\normphi$ corresponds to making some expected number of  swaps of adjacent candidates; of course, there is naturally some variance around this average. 
However, typically, for some fixed $\normphi$, for all swap distances with a non-negligible probability of getting sampled for this $\normphi$, in sampled elections at this distance the winning probabilities of candidates are typically quite similar (the only exception are very small values of $\normphi$). 
To further analyze the relationship between  $P_{E,c}(\normphi)$ and $Q_{E,c}(r)$, we computed the PCC of the $50\%$-winner threshold output by the two approaches on the synthetic dataset of \citet{DBLP:conf/atal/SzufaFSST20}.
For Plurality, Borda, Copeland, Bucklin, and STV, the correlation is $0.991$,  $0.982$, $0.987$, $0.989$, and $0.989$ respectively. 
So, overall, for both approaches the $50\%$-winner thresholds are very strongly correlated.

Finally, note that if we consider an election containing some top-truncated vote~$v$ (such votes  appear in our real-world data), then we do not adjust our procedure and still replace $v$ by a vote sampled from the Mallows distribution with $v$ as the central vote and the given normalized dispersion parameter.
This means that candidates that do not appear in the vote in the original election will never be added to it. 
We proceed in this way because otherwise we would need to make some (artificial) assumptions about the insertion probabilities of the non-ranked candidates. 
Moreover, in most of our applications, non-ranked candidates are not included in some vote ``by design''. For instance, in our political elections not all parties nominate a candidate in each voting district. 
Note also that in elections with top-truncated votes, winners can be both particularly robust and particularly non-robust: 
If the winner appears in all votes and all other candidates only appear in few votes, then the winner will still be the (by far) most probable winner for any value of $\normphi$, as even if each vote is replaced with a uniformly at random sampled one, the winner is most likely to have the strongest standing in the election. 
In contrast, top-truncated votes also open up the possibility for very non-robust winners: Consider as an example a Plurality election consisting of two candidates $c$ and $d$, where $x$ voters rank $c$ in the first position (and do not rank $d$ at all) and $x+1$ voters rank $d$ in the first position and $c$ in the second position. Then $d$ wins the election; however, in each election at swap distance $r>0$, $c$ wins.

\section{Comparing the Robustness of Different Voting Rules on Mallows Elections}  \label{sub:comp-mal}

 In this section, we conduct a comparison of the robustness of different voting rules using synthetic elections
generated from a variant of the Mallows model.

\paragraph{Setup}\label{subsub:setup}
For different voting rules, for $\normphi=\frac{1}{15}\cdot i$ with $i=0,\dots, 15$, we sampled $500$ elections with $10$ candidates and $100$ voters from the Mallows model with  lexicographic central order and normalized dispersion parameter $\normphi$. 
We reversed each sampled vote with probability $x\%$ for $x\in \{0,30,50\}$. 
The intuitive meaning of this model is that the electorate is split into two groups and the ``ground truth'' (the central vote in the Mallows model) of one group is the reversed ``ground truth'' of the other. 
Subsequently, for each sampled election $E$, for $\normphi=\{0,0.1,0.2,\dots,1\}$, we estimated $P_{E,c}(\normphi)$ using $500$ samples.\footnote{As reported by \citet{DBLP:conf/ijcai/BoehmerBFNS21} real-world elections are typically ``close'' to some elections generated from the Mallows model. We reverse votes with some probability because the resulting elections are more interesting from a robustness perspective, as they illustrate the different behavior of voting rules.} 

\paragraph{Results}
In \Cref{fig:diverse}, we compare the robustness of five voting rules.
The results draw a mixed picture: 
For reversion probability $0\%$ which means that we simply consider elections sampled from the Mallows model, all voting rules become less robust as $\normphi$ grows (which is also quite intuitive, as the votes in the sampled elections become more and more different from each other).
Moreover, there is a clear ranking of the voting rules in terms of their robustness independent of $\normphi$: 
Copeland and Borda produce the most robust results. 
Bucklin is the third most robust rule, then STV, and Plurality is the least robust rule.
The results highlight that rules are most robust if they take into account the ``full election'' without local focus, as this prevents the existence of strong ``hidden'' contenders.\footnote{An example of  such a ``hidden'' contender for both Bucklin and Plurality is the candidate $b$ from the election $E$ from the introduction (where there are $50$ votes with $a\succ b\succ \dots$ and $49$ votes with  $b\succ \dots \succ a$), as in this election candidate $b$ wins under both rules as soon as one of the first $50$ voters swaps $b$ and $a$. 
Moreover, due to the ``local'' nature of the rules, $a$ cannot easily gain additional points, as $a$ is ranked last in all votes where it is not ranked first, and for Plurality it only matters who is ranked in the first position and for Bucklin in this election it only matters who is ranked in one of the first two positions.
In contrast to this, under Borda, $a$ is also able to gain points if it is ranked in the last position
(so there exist single swaps by which $a$ can gain points which it has maybe lost by other swaps).}

Notably, the robustness difference between the rules is largest for $\normphi=1$: For Plurality and STV the average $50\%$-winner threshold here is around $0.2$, while for the other three rules it is around $0.4$. 
This large gap is quite remarkable, as these are in some sense the elections containing the least structure and information.

When we start to reverse the sampled votes with some probability, winners become less robust (which is quite intuitive, as in case we reverse half of the votes, in expectation the first and last candidate from the central vote are equally strong). 
For Plurality and STV for $\normphi\in [0,0.8]$, if we reverse votes with some probability, then the average $50\%$-winner threshold is simply shifted down by some constant value compared to the $50\%$-winner threshold if we do not reverse any votes.
In contrast to this, for the other three voting rules, the robustness of elections for $\normphi=0$ and $\normphi=1$ becomes more and more similar as we increase the reversion probability (see the dotted line in \Cref{fig:diverse} for reversion probability  $30\%$ and the dashed line for reversion probability  $50\%$).
For reversion probability $50\%$, for these rules, elections with $\normphi=1$ are even slightly more robust than the ones for $\normphi=0$.  

To explain this different behavior of the voting rules, let us focus for a moment on elections sampled from the Mallows model with $\normphi=0$, central vote $c_1\succ\dots\succ c_m$, and $50\%$ reversion probability: 
In these elections, around half of the votes, say $x$, are $c_1\succ\dots\succ c_m$ and the other half, say $y$, are $c_m\succ\dots\succ c_1$.
If $x>y$, then $c_1$ is the (strict) majority winner and thus the winner under Bucklin, Copeland, Plurality, and STV. 
It is also easy to see that $c_1$ is the Borda winner. 
However, the robustness of $c_1$ for the different rules varies substantially. 
For illustrative purposes, we just compare STV and Copeland. 
For STV, note that initially either $c_1$ or $c_m$ is ranked in the first position in every vote (and both appear roughly the same number of times in the first position). 
Thus, also after some swaps are performed, it is likely that $c_1$ and $c_m$ are the last two non-eliminated candidates when computing the STV winner.
Accordingly, the election boils down to a pairwise comparison between $c_1$ and $c_m$. 
For $c_m$ to win this comparison, it needs to be in front of $c_1$ in some votes where it was initially behind $c_1$. 
This requires $m-1$ specific swaps per vote, as in all such votes $v_1$ is initially ranked in the first and $c_m$ is ranked in the last position. 
In contrast to this, for Copeland, in the initial election $c_1$ has score $9$ and $c_2$ has score $7$ (because $c_2$ wins the pairwise comparison against all candidates except $c_1$). 
Thus, for $c_1$ to lose the election, $c_2$ only needs to win the pairwise comparison against $c_1$, which can be achieved by performing a single swap in $x-y$ of the votes where $c_1$ is ranked in the first and $c_2$ in the second position.  

Having explained the different behaviors of voting rules if we reverse votes, let us remark that from a normative perspective it seems to be more reasonable to expect that a winner in an election generated from the Mallows model with $\normphi=0$ and $50\%$ reversion probability is not too robust.
In the end, these elections will always only be decided by the (small) difference between the number of reversed and non-reversed sampled votes.
So from this perspective, Borda, Bucklin, and Copeland are advantageous here, despite (and in fact because) they are less robust. 

\section{Experiments on Real World Data}\label{se:real-world-exp}
We analyze the robustness of real-world election winners. 
For this, we not only use the original election data but also the original voting rule. 
We address the following four questions:
\begin{enumerate*}
 \item[\textbf{Q1.}] How sensitive are election winners to equiprobable noise swaps in different types of real-world elections?
 \item[\textbf{Q2.}] Are there real-world elections where very few random swaps change the election outcome with high probability? 
 \item[\textbf{Q3.}] Do the winning probabilities of candidates behave similarly in all ``close'' elections?
 \item[\textbf{Q4.}] Can the robustness of winners to random swaps be assessed via alternative (simpler) measures?
\end{enumerate*}

To answer these questions, we consider two types of real-world elections: sports elections  and political elections. 
In \Cref{se:Formula1} we analyze the robustness of winners of the Formula 1 World Championship.
After that, in \Cref{se:political}, we turn to high-stake political elections. 
As voters in large political elections typically do not reveal their full preferences, we focus on first-past-the-post elections where voters are partitioned into districts and each district sends one representative to the parliament, and we study the robustness of the party winning the most seats in such Plurality elections. 

To assess candidates' winning probabilities for different perturbation levels we used the same procedure as in the previous section. 
As described in \Cref{sec:assessing}, 
for each election $E=(C,V)$ we computed $P_{E,c}(\normphi)$ for $\normphi\in \{0,0.1,\dots, 1\}$. 
For each value of $\normphi$, we did so by sampling $500$ elections where each vote $v\in V$ is replaced by a vote drawn from $\mathcal{D}_{\text{Mallows}}^{v,\normphi}$ and recording for each candidate the fraction of these elections where it is a~winner.

\subsection{Formula 1} \label{se:Formula1}
We consider the $38$ editions of the Formula 1 World Championship between 1981 and 2018, where in each year, between $20$ to $47$ drivers competed in between $15$ to $21$ races.\footnote{In the corresponding election, we have one candidate for each driver and one voter for each race ranking the drivers according to the finishing position of the driver in this race. 
Drivers who did not participate in a race or did not finish it do not appear in the respective vote.
The elections were collected by \citet{extractingelections}.}
Each driver gets a certain number of points from each race depending on its finishing position, and the candidate with the highest number of points wins (this can be interpreted as applying a positional scoring rule to the corresponding election). 
Over the years, different scoring vectors were used.
We present them in \Cref{tab:f1-scores}.

\begin{table}[t]
    \centering
    \begin{tabular}{c|l}
        years &  \multicolumn{1}{c}{scoring vector}   \\\hline 
        2010-2018 & $\mathbf{s}_{2018}=(25,18,15,12,10,8,6,4,2,1,0,\dots, 0)$\\
        2003-2009 & $\mathbf{s}_{2009}=(10,8,6,5,4,3,2,1,0,\dots, 0)$\\
        1991-2002 & $\mathbf{s}_{2002}=(10,6,4,3,2,1,0,\dots, 0)$\\
        1981-1990 & $\mathbf{s}_{1990}=(9,6,4,3,2,1,0,\dots, 0)^{\dagger}$ \\
    \end{tabular}
    \caption{Scoring vectors used in different editions of the Formula 1 World Championship. $\dagger:$ Between 1981 and 1990, computing the final score of a candidate, instead of summing up its points from all races, only the $11$ highest scores were taken into account.} 
    \label{tab:f1-scores}
\end{table}

\subsubsection{General Overview of Results (\textbf{Q1}\&\textbf{Q2})} \label{sub:F1Q1}
\begin{figure*}[t]
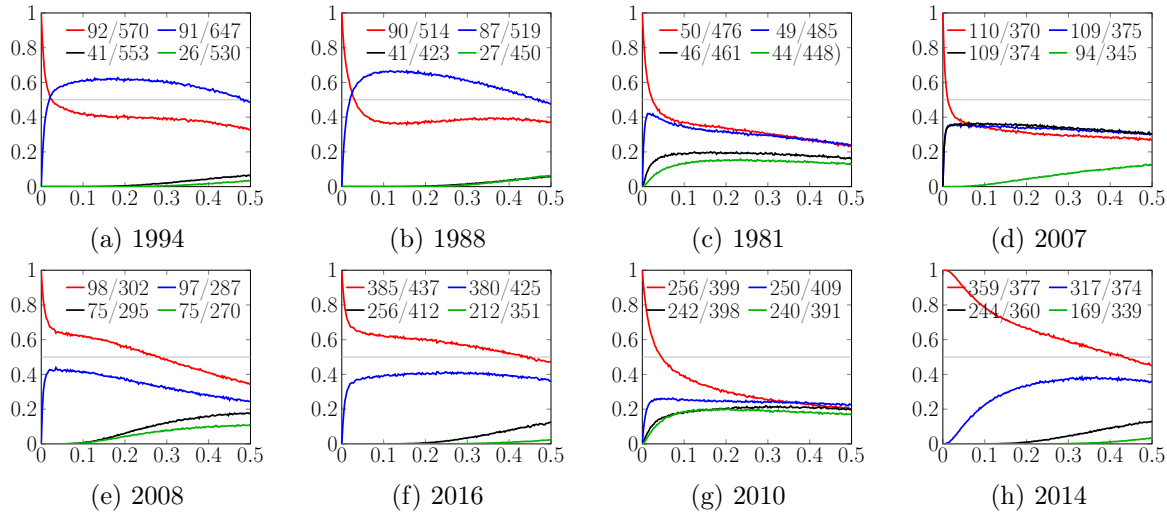

    \centering 
\begin{subfigure}{0.21\textwidth}
\hiddenforspeed{\resizebox{\textwidth}{!}{
}}
  \caption{1994}     
  \label{fig:Formula1}
\end{subfigure}\hfil 
\begin{subfigure}{0.21\textwidth}
  \hiddenforspeed{\resizebox{\textwidth}{!}{%
}}
  \caption{1988}
  \label{fig:Formula2}
\end{subfigure}\hfil 
\begin{subfigure}{0.21\textwidth}
  \hiddenforspeed{\resizebox{\textwidth}{!}{%
}}
  \caption{1981}
  \label{fig:Formula3}
\end{subfigure}\hfil
\begin{subfigure}{0.21\textwidth}\hiddenforspeed{\resizebox{\textwidth}{!}{%
}}
  \caption{2007}
  \label{fig:Formula4}
\end{subfigure}\hfil 
\begin{subfigure}{0.21\textwidth}
  \hiddenforspeed{\resizebox{\textwidth}{!}{%
}}
  \caption{2014} 
  \label{fig:Formula8} 
\end{subfigure}
\caption{Eight Formula 1 elections. We plot $P_{E,c}(\normphi)$ ($y$-axis) as a function of $\normphi$ ($x$-axis) for the
four most successful candidates. For each candidate, the first legend entry displays the score of the candidate in the election according to the used voting rule and the second entry displays its Borda score.}
\label{fig:Formula}
\end{figure*}

Generally speaking, it is surprising how fragile the victory of numerous Formula 1 world champions was: 
The average $50\%$-winner threshold in our dataset is only $0.36$\footnote{Remarkably, this means that the average $50\%$-winner threshold here is lower than the average $50\%$-winner threshold  of our considered rules on most of the Mallows elections analyzed in  \Cref{sub:comp-mal}. This is even more remarkable recalling that the voting rules used in Formula 1 elections are rather on the robust side, as each voter awards points to many different candidates.}, and eight elections have a $50\%$-winner threshold below $0.1$. 
Going into more detail, in \Cref{fig:Formula}, we visualize eight elections of special interest. 
\Cref{fig:Formula1,fig:Formula2,fig:Formula3,fig:Formula4,fig:Formula5,fig:Formula6} all display generally quite close election: In all six the losing probability
of the initial winner is already above $10\%$ at $\normphi=0.0025$. 
In $2007$, where $\normphi=0.0025$ corresponds to making on average $5$ \emph{random} swaps in the whole election, the losing probability of the original winner is even $22\%$. 
This is remarkable recalling that these elections are not artificial examples but come from the real world and recalling that we focus on random swaps, implying that there is also a very high chance that none of the top candidates are involved in a random swap, in which case the initial winner still wins.
In the $2007$ election, where the score of the red candidate is initially one higher than the score of the blue and the black candidate, a possible explanation for the observed general non-robustness is that the scoring vector  $\mathbf{s}_{2009}=(10,8,6,5,4,3,2,1,0,\dots, 0)$ was used. Thus, swapping down the red candidate in one vote or swapping up the black or blue candidate in one vote can suffice to make the red candidate lose the election, as a single swap can change scores by two.
In fact, taking a closer look, even in $20$ out of the $357$ elections at swap distance $1$ of this election the red candidate is not a winner. 
This means that even if we just make a single random swap the loosing probability of the red candidate is already $5.6\%$.  

\subsubsection{Different Types of Close Elections (\textbf{Q3})}\label{subsub:close}\label{sub:F1Q3}
While  all six elections from \Cref{fig:Formula1,fig:Formula2,fig:Formula3,fig:Formula4,fig:Formula5,fig:Formula6} were really close in the sense that few random swaps can have a crucial influence on the outcome, the (non)-robustness of the winners in these six elections still comes with quite different flavors.  
In \Cref{fig:Formula1,fig:Formula2}, the blue candidates overtakes the initially winning red candidate already at $\normphi=0.0275$ and afterwards consistently has a higher winning probability; in such elections one could say that the winner won more by luck or accident than by merit, as in most elections close to the original election a different candidate wins.
Let us focus for a moment on the 1994 Formula 1 World Championship with scoring vector $\mathbf{s}_{2002}=(10,6,4,3,2,1,0,\dots, 0)$, which consists of $16$ races and was decided by one point (\Cref{fig:Formula1}). 
What stands out is that the red candidate  won $8$ of the $16$ races and came in second in $2$ races, but either did not participate or did not complete the other $6$ races. 
Thus, if we perform a single random swap involving the red candidate, then in $8$ cases he loses $4$ points, in $2$ cases he loses $2$ points and in $2$ cases he gains $4$ points. 
Thus, for $10$ out of $12$ swaps involving the red candidate, the blue candidate wins the election after performing the swap (note, however, that only $7.4\%$ of all swaps involve one of the top-two candidates).
Accordingly, the general non-robustness of the red candidate here is due to the fact that the red candidate is much more likely to lose points instead of gaining more if few random swaps are performed.

In contrast to \Cref{fig:Formula1,fig:Formula2}, in \Cref{fig:Formula3,fig:Formula4}, the red candidate starts to have roughly the same winning probability as some other candidate(s) at small $\normphi$, however with increasing $\normphi$ the situation does not change. In such elections, it seems that the red candidate's victory was very fragile and that the top-two candidates are in fact of equal quality. 
Lastly, in \Cref{fig:Formula5,fig:Formula6}, while the red candidate already starts to have a significant loosing probability at small $\normphi$, its winning probability until $\normphi=0.5$ is always clearly the highest. 
In such elections it seems that the red candidates victory is a bit fragile but still ``justified'' and grounded on solid support.

\subsubsection{Relationship Between Winner Robustness and Candidate Scores (\textbf{Q4})} \label{sub:F1Q4}
Motivated by the observation that in all six considered close elections the score difference between the winner and the runner-up is between one and five and thus, in general, quite low, we now discuss the capabilities of the difference of the score of the election winner and runner-up to judge the robustness of winners. 
In general, there clearly is some correlation: 
The PCC of the score difference and the $50\%$-winner threshold in the Formula 1 elections is $0.66$ and, in particular, with increasing score difference, on average, winners get substantially more robust. 
However, the correlation is not strong and there also exist elections where there is a clear difference: 
For instance, in 2010 (\Cref{fig:Formula7}), the score difference is six but the $50\%$-winner threshold is still only $0.0475$ (and thus, in particular much lower than in 2016 (\Cref{fig:Formula6}) where the score difference is five). 
In 2014 (\Cref{fig:Formula8}), the score difference is $42$ and thus quite high, which is also reflected in a $50\%$-winner threshold of $0.4325$.
However, remarkably, the initial winner's loosing probability is already $1\%$ at $\normphi=0.0175$ and $10\%$ at $\normphi=0.04$, indicating that the election was much closer than what is suggested by the large score difference.
Further, note that in 1988 (\Cref{fig:Formula2}), where the red candidate seems to have won more by luck or accident than by merit, the score difference is three, whereas in 2008 (\Cref{fig:Formula5}), where the red candidate still dominates all other candidates in terms of winning probabilities even if many swaps are performed, the score difference is only one.

A different possible approach to identify (different types of) close elections could be to consider the candidates' Borda score.
The hope here is that the Borda score captures the general strength of the candidate in the election (which is not necessarily captured in the Formula 1 score, as here only points for finishing in one of the first positions are awarded). 
Generally speaking, the Borda score correlates with our classification of the six close elections from \Cref{fig:Formula1,fig:Formula2,fig:Formula3,fig:Formula4,fig:Formula5,fig:Formula6} (see \Cref{subsub:close}):
In particular, if the Borda score of one candidate is significantly higher than the scores of all other candidates, then this candidate will be the most probable winner for medium and large values of $\normphi$. 
However, also the Borda score has some clear limitations: 
In 1988 (\Cref{fig:Formula5}), the blue candidate has a lead of $4$ Borda points, while in 1981 (\Cref{fig:Formula3}) its lead is $9$ Borda points; nevertheless, in 1988 the blue candidate quickly becomes the most probable winner, which does not happen in 1981. 

\begin{figure*}[t]
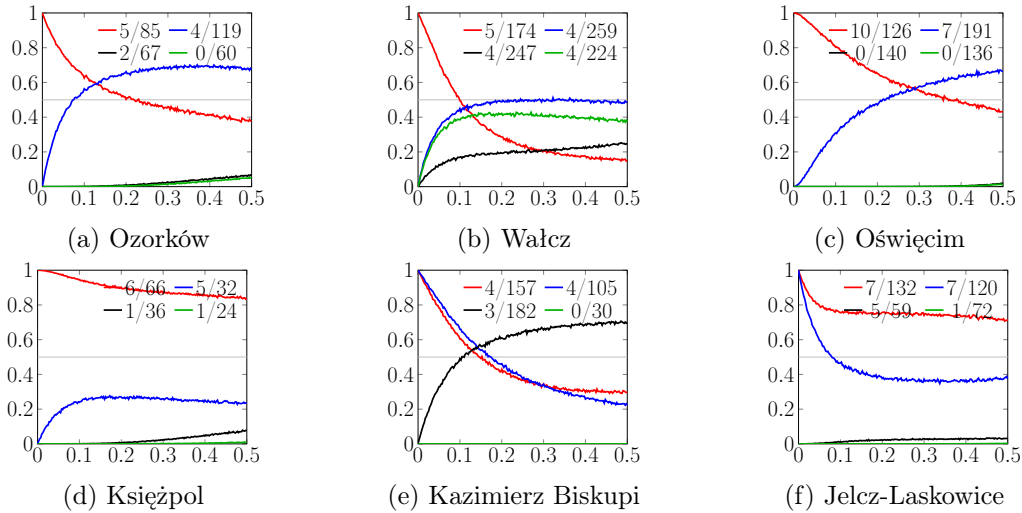

	\centering 
	\begin{subfigure}{0.21\textwidth}
		\hiddenforspeed{\resizebox{\textwidth}{!}{
}}
		\caption{Ozork\'{o}w}
		\label{fig:Poland1}
	\end{subfigure}\hfil 
	\begin{subfigure}{0.21\textwidth}
		\hiddenforspeed{\resizebox{\textwidth}{!}{%
}}
		\caption{Wa\l{}cz}
		\label{fig:Poland2}
	\end{subfigure}\hfil 
	\begin{subfigure}{0.21\textwidth}
		\hiddenforspeed{\resizebox{\textwidth}{!}{%
}}
		\caption{O\'{s}wi\k{e}cim}
		\label{fig:Poland3}
	\end{subfigure}

	\begin{subfigure}{0.21\textwidth}
		\hiddenforspeed{\resizebox{\textwidth}{!}{%
}}
		\caption{Ksi\k{e}\.{z}pol}
		\label{fig:Poland4}
	\end{subfigure} \hfil
	\begin{subfigure}{0.21\textwidth}
		\hiddenforspeed{\resizebox{\textwidth}{!}{%
}}
		\caption{Kazimierz Biskupi} 
		\label{fig:Poland5}
	\end{subfigure} \hfil
	\begin{subfigure}{0.21\textwidth}
		\hiddenforspeed{\resizebox{\textwidth}{!}{%
}}
		\caption{Jelcz-Laskowice}
		\label{fig:Poland6}
	\end{subfigure} 
	\caption{Six Poland elections. We plot $P_{E,c}(\normphi)$ as a function of $\normphi$ for the
		four most successful candidates. The legend displays the Plurality score and Borda score of each candidate in this order.}
	\label{fig:Poland}
\end{figure*}

\subsection{Political Elections} \label{se:political}

After we have seen that sports elections regularly have non-robust winners,  we now turn to a different type of election, political elections, mainly focusing on \textbf{Q1}\&\textbf{Q2}. 
As in large political elections full preferences of voters are typically unknown, we study a specific type of political election:
Several countries around the world use first-past-the-post voting in elections of different representative bodies.
In these elections, the voters are typically partitioned into constituencies, with each constituency having its distinct candidates.
Each constituency then sends the candidate with the highest number of votes to the representative body (and the representative body only consists of these candidates). 
We analyze the robustness of the strongest party in representative bodies elected by first-past-the-post elections.
For this, we identify each candidate running in some constituency by its party. 
Then, we create one vote for each constituency ranking in the $i$th position the party of the candidate finishing in the $i$th position in this constituency (in all elections we considered, there are never two candidates from the same party running in one constituency).
Now the Plurality score of a party in the constructed election is the number of seats the party gets in the representative body and the Plurality winner is the party with the highest number of seats.
Which party has won the most seats in a representative body is of great practical importance, as these parties are typically responsible for leading the formation of a new government and in some elections also decide on who should fill the most important political role (e.g., in UK general elections the winning party usually decides on who should be the Prime minister). 
Thus, if the robustness of the winning party is low, then one might want to consider a recount of the ballots or if the winner's robustness is low in some polls, then parties have additional motivation to mobilize as many voters as possible.

We observe that such political elections seem to be  very robust with the $50\%$-winner threshold typically being above $0.7$ and are thus in particular much more robust than the Formula 1 elections considered in \Cref{se:Formula1} and most of the synethetic elections analyzed in \Cref{sub:comp-mal}. 
Nevertheless, also non-robust winners regularly occur, highlighting the relevance of searching for them. 

\subsubsection{Polish Local Elections}
We analyze local council elections for different Polish cities from $2014$. 
In $2014$, in all cities with up to $\num{100000}$ inhabitants a first-past-the-post system was used. 
For this, all cities with up to $\num{20000}$/$\num{50000}$/$\num{100000}$ inhabitants where divided into $15$/$21$/$23$ constituencies. 
Our dataset consists of elections from $\num{1317}$ cities (we did not include elections with an average vote length
below $3$) each containing on average $8.6$ candidates.
Notably, out of the $\num{1317}$ elections $124$ are tied.
We generated the elections based on data provided by the Jagiellonian Center for Quantitative Research in Political Science.

Concerning the elections' robustness, in a large majority of the non-tied elections the winners are very robust; the average $50\%$-winner threshold is $0.78$ which is very high: In fact, only seven elections have a $50\%$-winner threshold below $0.2$ and only $98$ have a $50\%$-winner threshold below $0.5$. 
Overall, this is good news indicating that the considered type of political election is typically quite robust to random noise and that one does not have to worry about winners winning more by accident or luck than by merit.

Going into more details, in \Cref{fig:Poland}, six exemplary elections of special interest are shown. 
The two elections from Ozork\'{o}w (\Cref{fig:Poland1}) and Wa\l{}cz (\Cref{fig:Poland2}) were both decided by a single point (seat) and are in general quite close with a $50\%$-winner threshold of $0.22$ and $0.1$, respectively. 
Moreover, in both elections the initial winner has a $2\%$ losing probability already at $\normphi=0.0025$, which corresponds to making an expected number of around $0.37$ swaps in the full election. 
While it might sound counter-intuitive that a score difference of one can be overcome by making $0.37$ swaps, note that this is due to the fact that the Mallows distribution has some variance in the number of swaps that are performed.
To explain why winners have a non-negligible loosing probability already for small $\normphi$ in these elections, note that, for instance, in the election from Ozork\'{o}w (\Cref{fig:Poland1}) consisting of $10$ candidates and $15$ voters, four voters rank the red candidate in the first position and the blue candidate in the second position.
As there are overall $87$ different swaps, it follows that after making a single random swap the losing probability of the red candidate is $\frac{4}{87}=4.6\%$.  
This closeness highlights the importance of detecting such situations in order to be able to double check the integrity of the results. 

While such situations where already very few random swaps can change the election winner with a non-negligible probability occur mostly in elections with a score difference of one, there are also some (less) extreme examples with a larger score difference.
For instance, in O\'{s}wi\k{e}cim (\Cref{fig:Poland3}), the score difference is three but nevertheless, the loosing probability of the initial winner is already $1\%$ at $\normphi=0.015$ and $10\%$ at $\normphi=0.06$ (which corresponds to making an expected number of $2.3$, respectively, $9$ swaps in the whole election).
In contrast to the former three examples, there are also elections with very robust winners, even several with just a score difference of one. 
The election in Ksi\k{e}\.{z}pol (\Cref{fig:Poland4}) is an example; together with \Cref{fig:Poland1,fig:Poland2} this election also shows that only examining the Plurality scores is insufficient to judge the robustness of election winners. This general disconnect is also reflected in a low PCC of $0.48$ between the score difference and the $50\%$-winner threshold on the whole dataset.

Examining the $124$ tied elections, interestingly, our approach is able to identity different types of ties:
On the one hand, we have numerous tied elections where the winning probabilities of the different initially winning candidates behave very similarly if more and more swaps are performed (see \Cref{fig:Poland5} for an example; this election is also quite interesting because the initially third-ranked candidate seems to be particularly strong). 
On the other hand, in many of the tied elections, the winning probability of one of the winners decreases much faster than for the other, indicating that the later has a stronger general position in the election and is closer to uniquely winning the election than the other candidate (see \Cref{fig:Poland6} for an example).
This indicates that ties in elections might be of a different nature and that our approach might be a first possibility to better understand and identify them. 
In the analyzed political elections, tie-breaking is of special importance because usually one party is appointed to form a government. 

\subsubsection{UK General Elections} 
We now turn to national elections in the UK.
In particular, we consider the twelve general elections (of the House of Parliaments) in the UK which took place between 1974 and 2019. 
From this, we obtained twelve elections with between $9$ and $13$ candidates and $635$ and $659$ voters.
The elections were created by us based on data from the official website of the house of commons, \url{commonslibrary.parliament.uk}.

As in the Polish local elections, in general the robustness of winners is quite high in these elections; the average score difference between the winner and the runner-up is with $118$ also quite high.
In particular, only two out of the twelve analyzed elections have a $50\%$-winner threshold below $0.9$; both of them are from the year 1974. In this year, there was one election in February and a reelection in October: The October election has a $50\%$-winner threshold of $0.5$ and a score difference of $42$ (where only at $\normphi=0.1$ the winning probability of the initial winner drops below $99\%$). 
Thus, in this election, the winner is still pretty robust.
The February election is much closer. 
This election consists of $635$ votes over twelve candidates with an average vote length of $3.3$; the Labor party  won with $301$ seats against the Conservative party  with $297$ seats.
The $50\%$-winner threshold of this election is only $0.07$, which corresponds to performing $88.7$ swaps in the whole election in expectation. 
However, even for $\normphi=0.01$, which corresponds to making $12.67$ swaps in expectation, the losing probability of the Labor party is already $11.7\%$.\footnote{Notably, even at $\normphi=0.0025$, which corresponds to making an expected number of $3.2$ swaps in the election, the losing probability is already $1\%$ despite the fact that the initial score difference is $4$ (this is due to the fact that the Mallows model has some variance in the number of swaps it actually applies).}
The general non-robustness of the Labor parties victory in this election is also reflected in the candidates' Borda scores, as the Borda score of the Conservative party is $95$ points higher.
To sum up, UK general elections seem to be quite robust to our noise model; however, the February 1974 election constitutes a clear outlier as the win of the Labor party in this election is fragile. 
In fact, the Labor party did not win the absolute majority of seats in this election and, possibly as a consequence of the non-robustness of their victory, coalition talks failed. 
After the Labor party governed for a short time as a minority government, a reelection was initiated.   
In this reelection, the Labor party won again but this times with a larger lead (also being robust to random swaps). 
The clear difference between the two elections which happened in the span of $8$ months indicates that in political elections a significant fraction of voters can change their mind shortly after an election. 
This additionally motivates the study of the robustness of outcomes as an indicator for the likelihood that the outcome still reflects the voter's opinions even some time after the election, and also motivates that larger numbers of random swaps can realistically happen.

\section{Conclusion}
In this paper, we have studied how robust election winners are to equiprobable random noise by comparing different voting rules and computing and analyzing the robustness of real-world election winners. 
As one of our highlights, we have identified many real-world election winners that are very sensitive to random noise, indicating the these elections were extremely close. 
Moreover, we have illustrated that our approach can detect a variety of different patterns and can differentiate between seemingly very similar elections. 
For future work it would be interesting to dive deeper into the capabilities and limitations of our approach, for instance, by further exploring the possibility to use it as a tie-breaking mechanism. 

While we have already tried to make our experiments relevant to practitioners, there is certainly room for improvement: 
Because it is the (computationally) simpler and cleaner approach, we have considered an unweighted model where each swap has the same probability. 
However, this might not fully capture reality in all its facets: 
For instance, in the Formula 1 elections, one could argue that the probability of swapping two drivers in a race should be anti-proportional to their difference in finishing time.
Moving from the unweighted to the weighted setting would also require collecting the needed weighted data, which for some type of elections is also simply not available. 
From a data collection point of view, it would also be beneficial to collect the full preferences of voters in political elections (and not only top-choices as it is usually done) to analyze the robustness of large scale real-world political elections. 
Poll stations are probably the most promising starting point here. 
 
\subsection*{Acknowledgement}
We thank Nathan Schaar for helping us to collect the used real-world elections.
NB was supported by the DFG project ComSoc-MPMS (NI 369/22).
This project has received funding from the European 
    Research Council (ERC) under the European Union's Horizon 2020 
    research and innovation programme (grant agreement No 101002854).

\begin{center}
  \includegraphics[width=3cm]{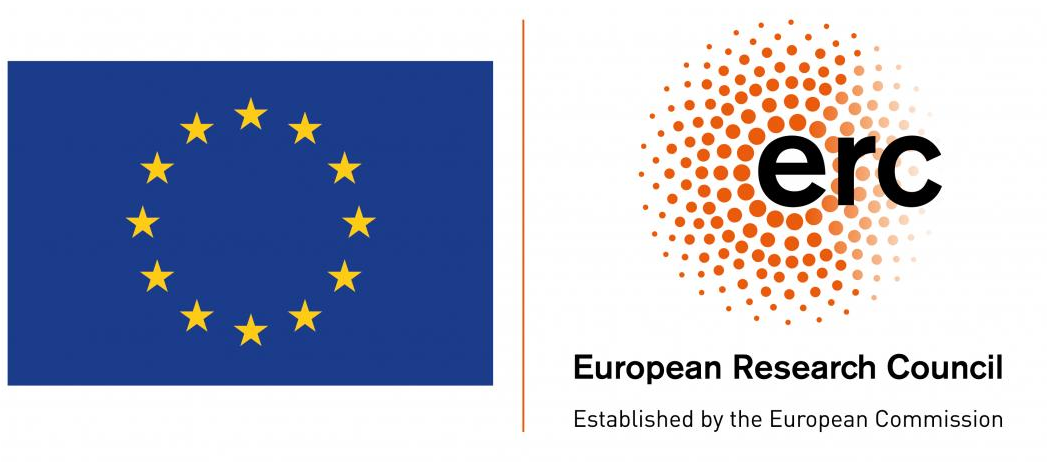}
\end{center}

\end{document}